\UseRawInputEncoding
\documentclass[11pt,nofootinbib,preprint]{revtex4}
\usepackage{mathrsfs}
\usepackage{graphicx}
\usepackage{amsmath}
\usepackage{amsfonts}
\usepackage{amssymb}
\usepackage{color}

\usepackage{epsfig}
\usepackage{CJK}
\usepackage{graphicx}
\usepackage{epsfig}
\usepackage{eepic}
\usepackage{bbm}
\usepackage{dcolumn}% Align table columns on decimal point
\usepackage{bm}% bold math
\usepackage{ulem}
\usepackage{slashbox}
\usepackage{multirow}% for the tabular, added in 30.01.2013
\usepackage{slashed}

\newcommand{\omits}[1]{}

\def\bc{\begin{center}}

\def\ec{\end{center}}
\def\be{\begin{eqnarray}}
\def\ee{\end{eqnarray}}

\definecolor{dyellow}{rgb}{1.,0.8,.0}
\definecolor{myblue}{rgb}{.1,.1,.7}
\definecolor{dcyan}{rgb}{.0,.6,.6}
\definecolor{cyan}{rgb}{0.4,1.0,1.0}
\definecolor{dmagenta}{rgb}{0.6,0.0,0.6}
\definecolor{brown}{rgb}{0.6,0.2,0.}
\definecolor{darkblue}{rgb}{.0,.0,0.5}
\definecolor{darkred}{rgb}{0.75,0.0,0.0}
\definecolor{orange}{rgb}{1.,.6,.0}
\definecolor{dorange}{rgb}{0.8,.4,.0}
\definecolor{green}{rgb}{0.0,1.0,0.0}
\definecolor{darkgreen}{rgb}{0.0,0.6,0.0}
\definecolor{purple}{rgb}{.4,.0,.4}
\definecolor{lightgrey}{rgb}{0.7, 0.7, 0.7}
\definecolor{grey}{rgb}{0.4, 0.4, 0.4}
\newcommand{\nc}{\newcommand}
\nc{\rnc}{\renewcommand} \nc{\ket}[1]{\left | \, #1 \right \rangle}
\nc{\bra}[1]{\left \langle #1 \, \right |}
\nc{\ua}{\uparrow} \nc{\da}{\downarrow}

\nc{\braket}[2]{\langle\, #1\,|\,#2\,\rangle}
\nc{\half}{\frac{1}{2}}

\nc{\prj}{\mathcal{P}} \nc{\hilb}{\mathcal{H}}
\nc{\pth}{\mathcal{C}} \nc{\inprod}[2]{\braket{#1}{#2}}
\nc{\upket}{\ket{\uparrow}} \nc{\downket}{\ket{\downarrow}}
\nc{\upbra}{\bra{\uparrow}} \nc{\downbra}{\bra{\downarrow}}

\begin{document}

%\preprint{hep-th/yymmnnn}

\title{Thread/State correspondence: from bit threads to qubit threads}

\author{Yi-Yu Lin$^{1,2}$} \email{yiyu@bimsa.cn}
\author{Jie-Chen Jin$^3$} \email{jinjch5@mail2.sysu.edu.cn}

\affiliation{${}^1$Beijing Institute of Mathematical Sciences and Applications (BIMSA),
	Beijing, 101408, China}
\affiliation{${}^2$Yau Mathematical Sciences Center (YMSC), Tsinghua University,
	Beijing, 100084, China}
\affiliation{${}^3$School of Physics and Astronomy, Sun Yat-Sen University, Guangzhou 510275, China}

\begin{abstract}

Starting from an interesting coincidence between the bit threads and SS (surface/state) correspondence, both of which are closely related to the holographic RT formula, we introduce a property of bit threads that has not been explicitly proposed before, which can be referred to as thread/state correspondence (see~\cite{Lin:2022agc} for a brief pre-release version). Using this thread/state correspondence, we can construct the explicit expressions for the SS states corresponding to a set of bulk extremal surfaces in the SS correspondence, and nicely characterize their entanglement structure. Based on this understanding, we use the locking bit thread configurations to construct a holographic qubit threads model as a new toy model of the holographic principle, and show that it is closely related to the holographic tensor networks, the kinematic space, and the connectivity of spacetime.

\end{abstract}

%% REVTEX4
\pacs{04.62.+v, 04.70.Dy, 12.20.-m}

\maketitle
\tableofcontents

%%%%%%%%%%%%%%%%%%%%%%%%%%%%%%%%%%%%%%%%%%%%%%%%%%%%%%%%%%%%%%%%%%%%%%
\section{Introduction}
%%%%%%%%%%%%%%%%%%%%%%%%%%%%%%%%%%%%%%%%%%%%%%%%%%%%%%%%%%%%%%%%%%%%%%

Advances in recent decades suggest that we may have found an important and marvelous clue or ``belief'' to the mystery between gravity and quantum mechanics, namely ``It from Qubit"~\cite{VanRaamsdonk:2010pw,Lashkari:2013koa,Faulkner:2013ica,Faulkner:2017tkh}. In this view, spacetime is not a fundamental object, but rather emerges from a structure of quantum entanglement. The clue is successively related to the concepts of the black hole area entropy~\cite{Bekenstein:1972tm,Bekenstein:1973ur,Bekenstein:1974ax,Bardeen:1973gs}, the holographic principle (especially AdS/CFT duality)~\cite{Maldacena:1997re,Gubser:1998bc,Witten:1998qj}, and the RT formula of the holographic entanglement entropy~\cite{Ryu:2006bv,Ryu:2006ef,Hubeny:2007xt}, which have built a bridge between quantum mechanics and general relativity. In particular, the RT formula shows that the entanglement entropy that characterizes quantum entanglement between different parts of a particular class of (i.e., ``holographic") quantum systems can be equivalently (i.e., ``dually") described by the area of an extremal surface in a corresponding curved spacetime~\cite{Ryu:2006bv,Ryu:2006ef,Hubeny:2007xt}.

More recently, many enlightening ideas from other fields, such as condensed matter physics, quantum information theory, network flow optimization theory, etc., have entered and benefited the study of the holographic gravity. One of the most striking examples is that inspired by the tensor network method originally used in condensed matter physics as a numerical simulation tool to investigate the wave functions of quantum many-body systems, various holographic tensor network (TN) models have been constructed as toy models of the holographic duality, such as MERA (multiscale entanglement renormalization ansatz) tensor network~\cite{Vidal:2007hda,Vidal:2008zz,Swingle:2009bg,Swingle:2012wq}, perfect tensor network~\cite{Pastawski:2015qua}, random tensor network~\cite{Hayden:2016cfa}, p-adic tensor network~\cite{Chen:2021ipv,Chen:2021qah,Chen:2021rsy}, OSED (one-shot entanglement distillation) tensor network~\cite{Bao:2018pvs,Bao:2019fpq,Lin:2020ufd} and so on. For more research on tensor networks in the holographic context, see e.g.~\cite{ Hung:2019zsk, Milsted:2018vop,Milsted:2018yur, Milsted:2018san, SinaiKunkolienkar:2016lgg,Bao:2017qmt,Beny:2011vh, Qi:2013caa,Ling:2019akz,Ling:2018ajv,Ling:2018vza,Bhattacharyya:2017aly,Bhattacharyya:2016hbx, Gan:2017nyt,Bao:2015uaa,Yu:2020zwk,Sun:2019ycv}. Furthermore, inspired by the continuous version of the holographic tensor network models, especially the construction of cMERA (continuous MERA) tensor network~\cite{Haegeman:2011uy}, ~\cite{Miyaji:2015fia,Miyaji:2015yva} proposed the so-called SS correspondence (surface/state correspondence) as a more specific mechanism of the holographic principle. The SS duality refers to the duality between a codimension two convex surface $\Sigma $ in the holographic bulk spacetime and a quantum state described by a density matrix $\rho (\Sigma )$, which is defined on the Hilbert space of the quantum theory dual to the Einstein's gravity. This can be understood intuitively in the context of tensor networks. For a convex closed surface $\Sigma $, we can always contract the indices of tensors contained in the region enveloped by $\Sigma $ to obtain a state $\rho (\Sigma )$.

Another idea to further explore the profound connection between spacetime geometry and quantum entanglement is inspired by the optimization problem in network flow theory.~\cite{Freedman:2016zud,Cui:2018dyq,Headrick:2017ucz} developed the optimization theory of flows on manifolds, endowed with the name of ``bit threads", and proposed the concept of bit threads can equivalently formulate the RT formula. Although bit threads are visually intuitive and usually interpreted implicitly as the bell pairs distilled from the boundary quantum systems, they are mostly used merely as a mathematical tool to study different aspects of holographic principles (for the recent developments of bit threads see e.g.~\cite{Lin:2021hqs,Lin:2020yzf,Headrick:2020gyq,Agon:2021tia,Rolph:2021hgz,Chen:2018ywy,Hubeny:2018bri,Agon:2018lwq,Du:2019emy,Bao:2019wcf,Harper:2019lff,Agon:2019qgh,Du:2019vwh,Agon:2020mvu,Bao:2020uku,Pedraza:2021fgp,Pedraza:2021mkh,Harper:2018sdd,Shaghoulian:2022fop,Susskind:2021esx,Lin:2022agc,Headrick:2022nbe,Harper:2021uuq,Harper:2022sky,Lin:2022aqf,Kudler-Flam:2019oru}). Among these developments, it is worth noting that in~\cite{Lin:2020yzf} bit threads are related with the holographic entanglement distillation tensor networks, and in~\cite{Lin:2021hqs}, the component flow fluxes in the locking bit thread configurations are shown to explain the partial entanglement entropies of the boundary  quantum systems. In this paper~\footnote{For a brief pre-release version containing the core of this work, see~\cite{Lin:2022agc}.}, by studying the connection between bit threads and SS duality, we propose a natural and novel physical property of bit threads, dubbed ``thread/state correspondence'', that is, in a so-called locking bit thread configuration~\cite{Headrick:2020gyq,Lin:2020yzf,Lin:2021hqs,Lin:2022aqf}, each bit thread is in a quantum superposition state of two orthogonal states. Our thread/state rules explicitly accentuate for the first time the implied meaning of the name ``bit threads'', namely, that these threads, which are used mathematically to recover the RT formula, can be physically assigned a meaning closely related to the concept of ``bits''. Moreover, since the fluxes of bit threads are directly related to the geometric quantities of a holographic spacetime, while this newly proposed thread-state property can cleverly characterize quantum entanglement as one will see, it can be expected to be a very useful advance in the study of the relationship between quantum entanglement and spacetime geometry. 

More explicitly, this paper will show that, using thread/state rules, we can do (but expect to do more than) the following things: first we can construct the explicit expressions for the SS states corresponding to a set of bulk extremal surfaces in the SS duality, and nicely characterize their entanglement structure; use the locking bit thread configurations to construct a new toy model of the holographic principle and actually, we explicitly give the relationship between this model and the holographic tensor network models; naturally understand the so-called kinematic space~\cite{Czech:2015kbp,Czech:2015qta}, endow it with the interpretation of microscopic states such that to explain that the entropy is proportional to volume therein; in some sense quantitatively characterize the famous ``It from qubit'' thought experiment~\cite{VanRaamsdonk:2010pw}, that is, by removing the entanglement in the boundary quantum system, the bulk spacetime will accordingly deform, or even break up.

The idea of thread-state will provide a complementary perspective distinct from the local tensors used in holographic TN models. Moreover, our thread/state prescription for locking thread configurations is an enlightening step towards the issue of spacetime emergence. It is intriguing to find the similar rules for the more general non-locking bit thread configurations, then it is possible to further read the SS states of the general bulk surfaces. It is even more tantalizing to completely reconstruct the bulk geometry merely from the properties of the quantum state assigned to the bit threads. In addition, it is also interesting to consider how the thread/state rules adapt to the covariant bit threads~\cite{Headrick:2022nbe} of the covariant RT formula~\cite{Hubeny:2007xt}, the quantum bit threads~\cite{Agon:2021tia,Rolph:2021hgz} that can account for the bulk quantum corrections to the RT formula~\cite{Faulkner:2013ana,Engelhardt:2014gca}, the Lorentzian bit threads~\cite{Pedraza:2021fgp,Pedraza:2021mkh} that can characterize the holographic complexity~\cite{Brown:2015bva,Susskind:2014rva}, the hyperthreads~\cite{Harper:2021uuq,Harper:2022sky} that can study the multipartite entanglement, and so on, and may provide useful insights on all of these topics.

The structure of this paper is as follows: Section~\ref{sec2} is the background review of the basic knowledge of bit threads and the surface/state correspondence. Section~\ref{sec3} presents the motivation for this work. In Section~\ref{subsec3.1} we study the problem of constructing SS states of bulk extremal surfaces, then in Section~\ref{subsec3.2} we propose the thread/state correspondence as a natural and efficient prescription for this problem. Based on the interesting connection between bit threads and SS correspondence, in Section~\ref{sec4} we construct a holographic qubit threads model using locking bit thread configurations, wherein the bit threads can be understood as qubit threads or CMI threads. In Section~\ref{subsec5.1} and~\ref{subsec5.2}, we discuss its close connection with the holographic tensor network model and the holographic kinematic space respectively. In Section~\ref{sec6} we use our qubit threads model to discuss the connectivity of spacetime, and show that qubit threads can play the roles of ``sewing'' a spacetime in a sense.

%%%%%%%%%%%%%%%%%%%%%%%%%%%%%%%%%%%%%%%%%%%%%%%%%%%%%%%%%%%%%%%%%%%%%%

%%%%%%%%%%%%%%%%%%%%%%%%%%%%%%%%%%%%%%%%%%%%%%%%%%%%%%%%%%%%%%%%%%%%%%
\section{Background review }\label{sec2}

This section is a brief review of bit thread, locking bit thread configuration and surface/state duality. Readers who are familiar with these aspects may skip to section~\ref{sec3}.
%%%%%%%%%%%%%%%%%%%%%%%%%%%%%%%%%%%%%%%%%%%%%%%%%%%%%%%%%%%%%%%%%%%%%%
%%%%%%%%%%%%%%%%%%%%%%%%%%%%%%%%%%%%%%%%%%%%%%%%%%%%%%%%%%%%%%%%%%%%%%
%%%%%%%%%%%%%%%%%%%%%%%%%%%%%%%%%%%%%%%%%%%%%%%%%%%%%%%%%%%%%%%%%%%%%%

\subsection{Bit threads and locking bit thread configurations }\label{subsec2.1}

\begin{figure}[htbp]     \begin{center}
		\includegraphics[height=7cm,clip]{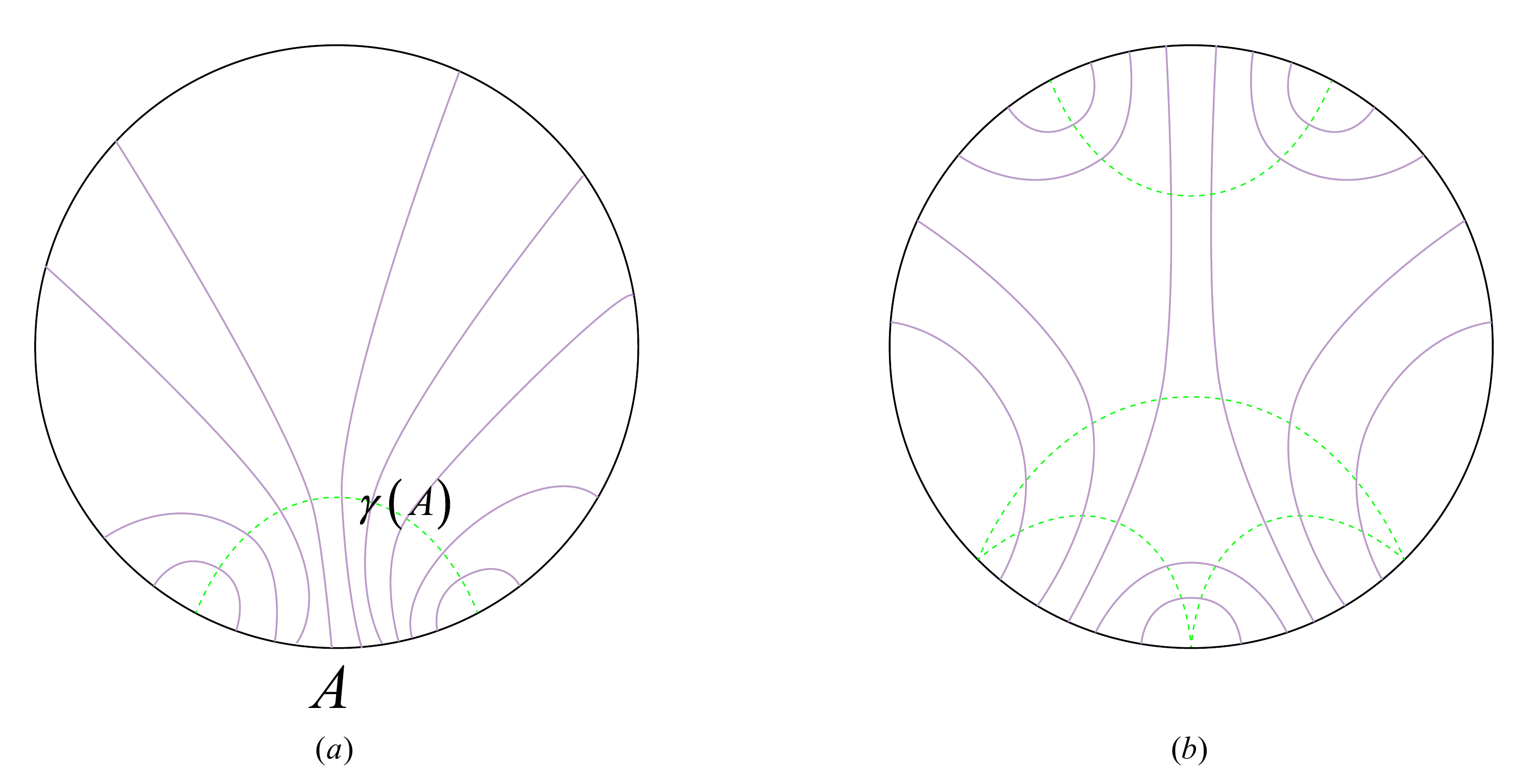}
		\caption{Two typical examples of locking bit thread configurations, where the mauve lines represent bit threads and the green dashed lines represent RT surfaces. (a) A simplest bit thread configuration locking $\left\{ {A,{A_c}} \right\}$ simultaneously, from which $S\left( A \right) = Flu{x_{locking}}\left( A \right)$.  (b) A locking bit thread configuration that can lock a series of non-crossing boundary regions simultaneously.  
		}
		\label{fig1}
	\end{center}	
\end{figure}
In the framework of holographic principle, bit threads are unoriented bulk curves which end on the boundary and subject to the rule that the thread density is less than $1/4{G_N}$ everywhere~\cite{Freedman:2016zud,Cui:2018dyq,Headrick:2017ucz}.~\footnote{When  one  takes  the  Hodge  dual  of  bit  threads  one  gets  calibrated  geometries,  which  mathematicians (geometers)  use  to  identify  minimal  surfaces.   This  is  a  viewpoint  adopted  in~\cite{Bakhmatov:2017ihw}.}   Mathematically, this thread density bound implies that the number of threads passing through the minimal surface $\gamma \left( A \right)$ that separates a boundary subregion $A$ and its complement ${{A_{\rm{c}}}}$ cannot exceed its area $Area\left( {\gamma \left( A \right)} \right)$, hence the flux of bit threads $Flux\left( A \right)$ connecting $A$ and its complement ${{A_{\rm{c}}}}$ does not exceed $Area\left( {\gamma \left( A \right)} \right)$:
\be\label{bound} Flux\left( A \right) \le \frac{1}{{4{G_N}}} \cdot Area\left( {\gamma \left( A \right)} \right).\ee
Borrowing terminology from the theory of flows on networks, a thread configuration is said to $lock$ the region $A$ when the bound~(\ref{bound}) is saturated. Actually, this bound is tight: for any $A$, there does exist a locking thread configuration satisfying:
\be Flux\left( A \right) = \frac{1}{{4{G_N}}} \cdot Area\left( {\gamma \left( A \right)} \right).\ee
This theorem is known as max flow-min cut theorem (see \cite{Headrick:2017ucz} and references therein), that is, the maximal flux of bit threads (over all possible bit thread configurations) connecting a boundary subregion $A$ of a Riemannian manifold with its complement ${{A_{\rm{c}}}}$  is equal to the area of the bulk minimal surface $\gamma \left( A \right)$ homologous to $A$. Therefore, the famous RT formula which relates the entanglement entropy of a boundary subregion $A$ and the area of the bulk minimal extremal surface ${\gamma \left( A \right)}$ homologous to A~\cite{Ryu:2006bv,Ryu:2006ef,Hubeny:2007xt}:
\be S\left( A \right) = \frac{{Area\left( {\gamma \left( A \right)} \right)}}{{4{G_N}}},\ee
can be expressed in another way, that is, the entropy of a boundary subregion $A$ is equal to the flux of the threads connecting $A$ with ${{A_{\rm{c}}}}$ in a locking thread configuration:
\be S\left( A \right) = Flu{x_{locking}}\left( A \right).\ee
See figure~\ref{fig1}(a).

When the bit threads are required to be locally parallel, one can use the language of $flow$ to describe the behavior of bit threads conveniently in mathematics, that is, using a vector field $\vec v$ to describe the bit threads, just as using the magnetic field $\vec B$ to describe the magnetic field lines. The difference is that for the latter we regard the magnetic field itself as the more fundamental concept, while for the former we consider the threads to be more fundamental. The constraints on the bit threads can then be expressed as the requirements for the flow $\vec v$ as follows,
\be\nabla  \cdot \vec v &=& 0,\\
\rho \left( {\vec v} \right) &\equiv & \left| {\vec v} \right| \le 1/4{G_N}.\ee
For situations involving more than one pair of boundary subregions, the concept of $thread~ bundles$ is also useful. The threads in each thread bundle are required to connect only a specified pair of boundary subregions, while still satisfy the constraints of bit threads. Specifically, one can use a set of vector fields ${\vec v_{ij}}$ to represent each thread bundle connecting the ${A_i}$ region and ${A_j}$ region respectively. The set $V$ of vector fields ${\vec v_{ij}}$ is referred to as a $multiflow$, and each ${\vec v_{ij}}$ is called a $component~ flow$.

For a single boundary subregion $A$, the max flow-min cut theorem directly indicates that one can find a thread configuration that can lock the specified boundary subregion (and its complement simultaneously). In other words, there exist thread configurations that can lock a set of boundary regions $I = \left\{ {{A_1},{A_{\rm{c}}}} \right\}$, and there is typically an infinite number of choices. However, one can further ask, can we find a locking thread configuration that can lock an arbitrary specified set of subregions simultaneously? The question becomes very nontrivial. Broadly speaking, it depends not only on the relative space position relations between these specified subregions, but also on the properties we assign to the bit threads, in particular, the precise definition of the thread density bound. Recently, the authors in~\cite{Headrick:2020gyq} investigated this issue in great detail. They proposed and proved several theorems on the existence of locking thread configurations in various situations. Here, we merely point out one interesting locking theorem. See figure~\ref{fig1}(b) (For more existence theorems and detailed technical proofs of the theorems, see the original paper~\cite{Headrick:2020gyq}):

{\bf Non-crossing locking theorem:} There exists a multiflow that can lock all the elementary regions and all non-crossing composite regions simultaneously.

In this statement, we have divided (the time slice of) the holographic boundary system $\partial M$ into adjacent non-overlapping subregions ${A_1}, \ldots ,{A_n}$, which are referred to as $elementary~regions$, satisfying ${A_i} \cap {A_j} = \emptyset $, $\mathop  \cup \limits_{i = 1}^n {A_i} = \partial M$. Accordingly, a $composite~ region$ is defined as the union of some certain elementary regions. Furthermore, we are following the terminology from network theory: two boundary regions are said to cross if they partially overlap and do not cover the whole boundary. For example, $AB$ crosses $BC$, but does not cross $A$, $ABC$, or $D$. More explicitly, two regions $X$ and $Y$ do not cross if and only if at least one of the following conditions holds:
$ X \cap Y = \emptyset ,\quad X \subseteq Y,\quad Y \subseteq X,\quad X \cup Y = \partial M $.

\begin{figure}[htbp]     \begin{center}
		\includegraphics[height=4.5cm,clip]{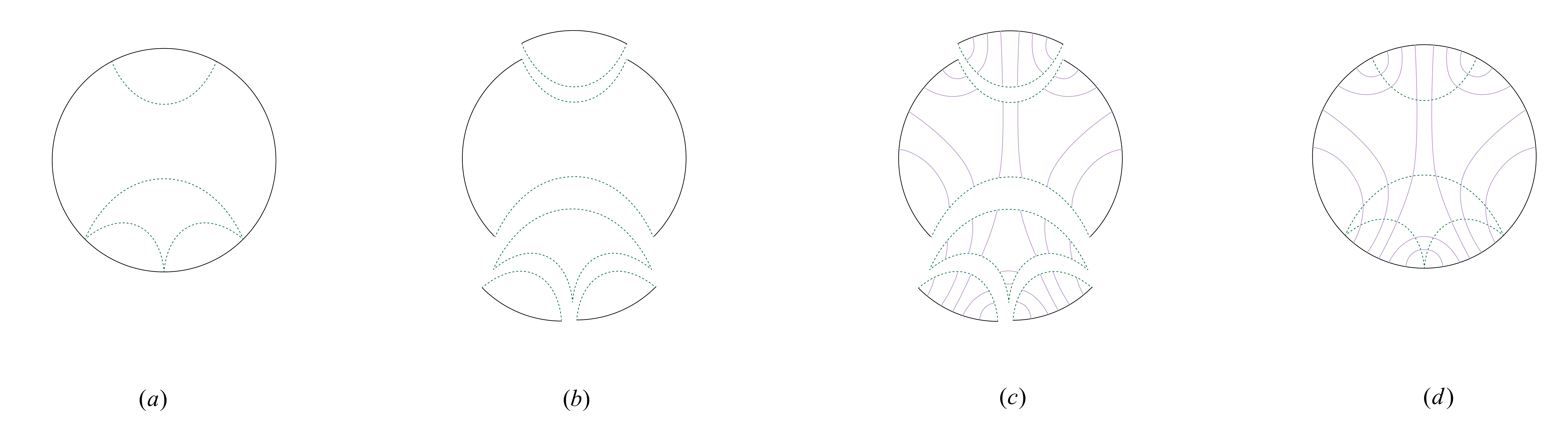}
		\caption{Bulk-cell decomposition method (where the mauve lines represent bit threads and the green dashed lines represent RT surfaces): (a) A specified set of non-crossing boundary regions corresponds to a set of non-intersecting RT surfaces; (b) Utilize the series of RT surfaces to discretize the bulk spacetime; (c) Regarding each cell as an individual spacetime, one can always find a locking thread configuration to lock all boundaries of each cell spacetime simultaneously; (d) Finally, by gluing all the locking thread configurations of cell spacetimes together, one can obtain a locking thread configuration of a whole bulk spacetime, and thus prove the existence.
		}
		\label{fig2}
	\end{center}	
\end{figure}

This non-crossing locking theorem has a clever proof, which can be called ``bulk-cell decomposition'' method or ``bulk-cell gluing'' method ~\cite{Headrick:2020gyq}. This intuitive proof method can be described as follows: as shown in figure~\ref{fig2}, considering a specified set of boundary elementary regions and some non-crossing composite regions, it is easy to see that, since these regions do not cross each other, accordingly, the corresponding RT surfaces of these regions will not intersect with each other. Therefore, we can utilize this series of RT surfaces to discretize the bulk spacetime, according to some certain appropriate order. These small fragments cut by RT surfaces can be called ``cells'', and each cell is bounded by RT surfaces or boundary subregions. Next, regarding each cell as an individual spacetime, by using a simpler locking theorem, which states that there is always a locking bit thread configuration can lock all elementary regions simultaneously, one can always find a locking thread configuration to lock all boundaries of each cell spacetime simultaneously. Since the boundaries of the cell spacetime are minimal surfaces, the locking thread configuration must satisfy that the threads intersecting these surfaces should be orthogonal to them and saturate the density bound. Finally, by gluing all the locking thread configurations of cell spacetimes together, one can obtain a locking thread configuration of a whole bulk spacetime, and thus prove the existence.

%%%%%%%%%%%%%%%%%%%%%%%%%%%%%%%%%%%%%%%%%%%%%%%%%%%%%%%%%%%%%%%%%%%%%%
\subsection{Surface/state correspondence}\label{subsec2.2}
%%%%%%%%%%%%%%%%%%%%%%%%%%%%%%%%%%%%%%%%%%%%%%%%%%%%%%%%%%%%%%%%%%%%%%
%%%%%%%%%%%%%%%%%%%%%%%%%%%%%%%%%%%%%%%%%%%%%%%%%%%%%%%%%%%%%%%%%%%%%%
\begin{figure}[htbp]     \begin{center}
		\includegraphics[height=9cm,clip]{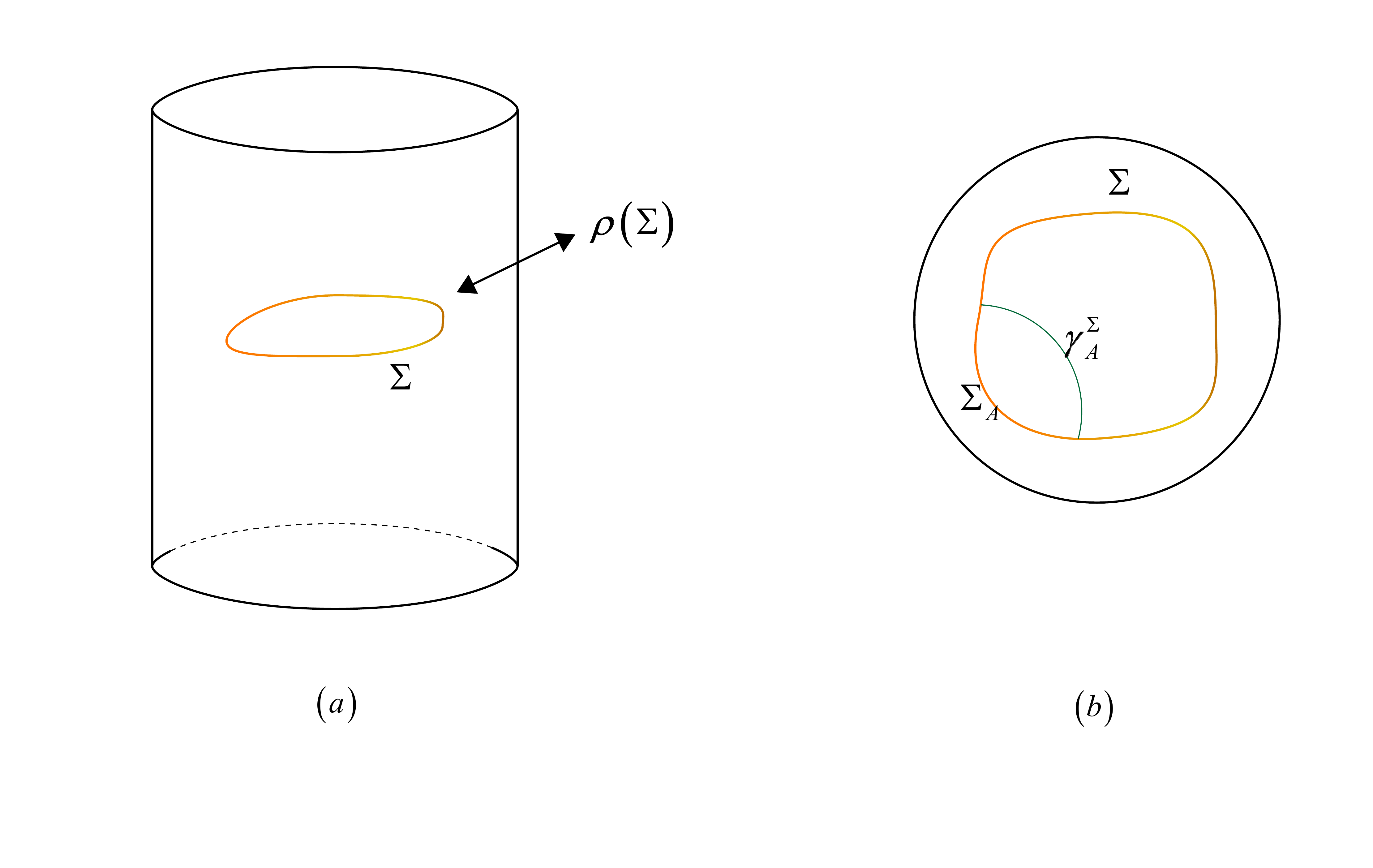}
		\caption{	(a)	Surface/state correspondence. (b) Generalized RT formula.
		}
		\label{fig3}
	\end{center}	
\end{figure}
Inspired by the construction of cMERA~\cite{Haegeman:2011uy}, a continuous version of the MERA tensor network, ~\cite{Miyaji:2015yva,Miyaji:2015fia} proposed the so-called surface/state correspondence as a more concrete mechanism of the holographic principle. In principle, this duality can be applied to any spacetime described by Einstein's gravity, even those without timelike boundaries. Surface/state correspondence (or SS duality) refers to the duality between a codimension two convex surface $\Sigma $ in the holographic bulk spacetime and a quantum state described by a density matrix $\rho (\Sigma )$, which is defined on the Hilbert space of the quantum theory dual to the Einstein's gravity.

In SS duality, when a surface is closed and topologically trivial, its dual quantum state (we will name it as SS state) is given by a pure state $\rho \left( \Sigma  \right) = \left| \Sigma  \right\rangle \left\langle \Sigma  \right|$. In particular, if we consider an Einstein’s gravity in an AdS space and take $\Sigma $ to be a time slice of the AdS boundary, then $\left| {\Phi \left( \Sigma  \right)} \right\rangle $ is simply given by the ground state $\left| 0 \right\rangle $ of the dual CFT. Another interesting example is the SS state dual to a closed surface of zero size (i.e., a point). Such a state is identified as the so-called boundary state~$\left| B \right\rangle $~\cite{Miyaji:2014mca}. This is because according to the idea of holographic entanglement entropy, the states dual to such point-like surfaces have no real space entanglement, while the boundary states can characterize this feature.

More generally, the surface/state correspondence claims that, as shown in figure~\ref{fig3}, if we consider an arbitrary closed convex surface $\Sigma$ in the bulk spacetime and a subregion ${\Sigma _A}$ of $\Sigma$, then this closed surface $\Sigma$ and the surface ${\Sigma _A}$ will correspond to quantum states described by density matrices $\rho \left( \Sigma  \right)$ and $\rho \left( {{\Sigma _A}} \right)$ respectively, and the entanglement entropy $S_A^\Sigma $ of subregion ${\Sigma _A}$ with respect to the quantum state $\rho \left( \Sigma  \right)$, i.e., the von Neumann entropy of $\rho \left( {{\Sigma _A}} \right)$ can be calculated by the area of the extremal surface $\gamma _A^\Sigma $ anchored on the boundary of ${\Sigma _A}$, i.e.
\be\label{gene0} S_A^\Sigma =\frac{{\rm Area}\left( {\gamma _A^\Sigma } \right)}{4G}.
\ee
This can be viewed as a generalized version of RT formula, which suggests that the extremal surfaces in any bulk regions have definite physical meanings. As a result, in the surface/state correspondence, there is a general rule for the extremal surfaces as follows \cite{Miyaji:2015yva}: The density matrix corresponding to an extremal surface is a direct product of density matrices at each point, which means that the von Neumann entropy of the density matrix of an extremal surface is equal to its area (or in the language of tensor networks, it is just proportional to the number of bonds passing through the surface ).

%%%%%%%%%%%%%%%%%%%%%%%%%%%%%%%%%%%%%%%%%%%%%%%%%%%%%%%%%%%%%%%%%%%%%%

%%%%%%%%%%%%%%%%%%%%%%%%%%%%%%%%%%%%%%%%%%%%%%%%%%%%%%%%%%%%%%%%%%%%%%
\section{Motivation and thread/state prescription }\label{sec3}
%%%%%%%%%%%%%%%%%%%%%%%%%%%%%%%%%%%%%%%%%%%%%%%%%%%%%%%%%%%%%%%%%%%%%%
%%%%%%%%%%%%%%%%%%%%%%%%%%%%%%%%%%%%%%%%%%%%%%%%%%%%%%%%%%%%%%%%%%%%%%
%%%%%%%%%%%%%%%%%%%%%%%%%%%%%%%%%%%%%%%%%%%%%%%%%%%%%%%%%%%%%%%%%%%%%%

%%%%%%%%%%%%%%%%%%%%%%%%%%%%%%%%%%%%%%%%%%%%%%%%%%%%%%%%%%%%%%%%%%%%%%
\subsection{Motivation: constructing the SS states of bulk extremal surfaces by ``SS bit model'' }\label{subsec3.1}
%%%%%%%%%%%%%%%%%%%%%%%%%%%%%%%%%%%%%%%%%%%%%%%%%%%%%%%%%%%%%%%%%%%%%%
%%%%%%%%%%%%%%%%%%%%%%%%%%%%%%%%%%%%%%%%%%%%%%%%%%%%%%%%%%%%%%%%%%%%%%
%%%%%%%%%%%%%%%%%%%%%%%%%%%%%%%%%%%%%%%%%%%%%%%%%%%%%%%%%%%%%%%%%%%%%%
\begin{figure}[htbp]     \begin{center}
		\includegraphics[height=7cm,clip]{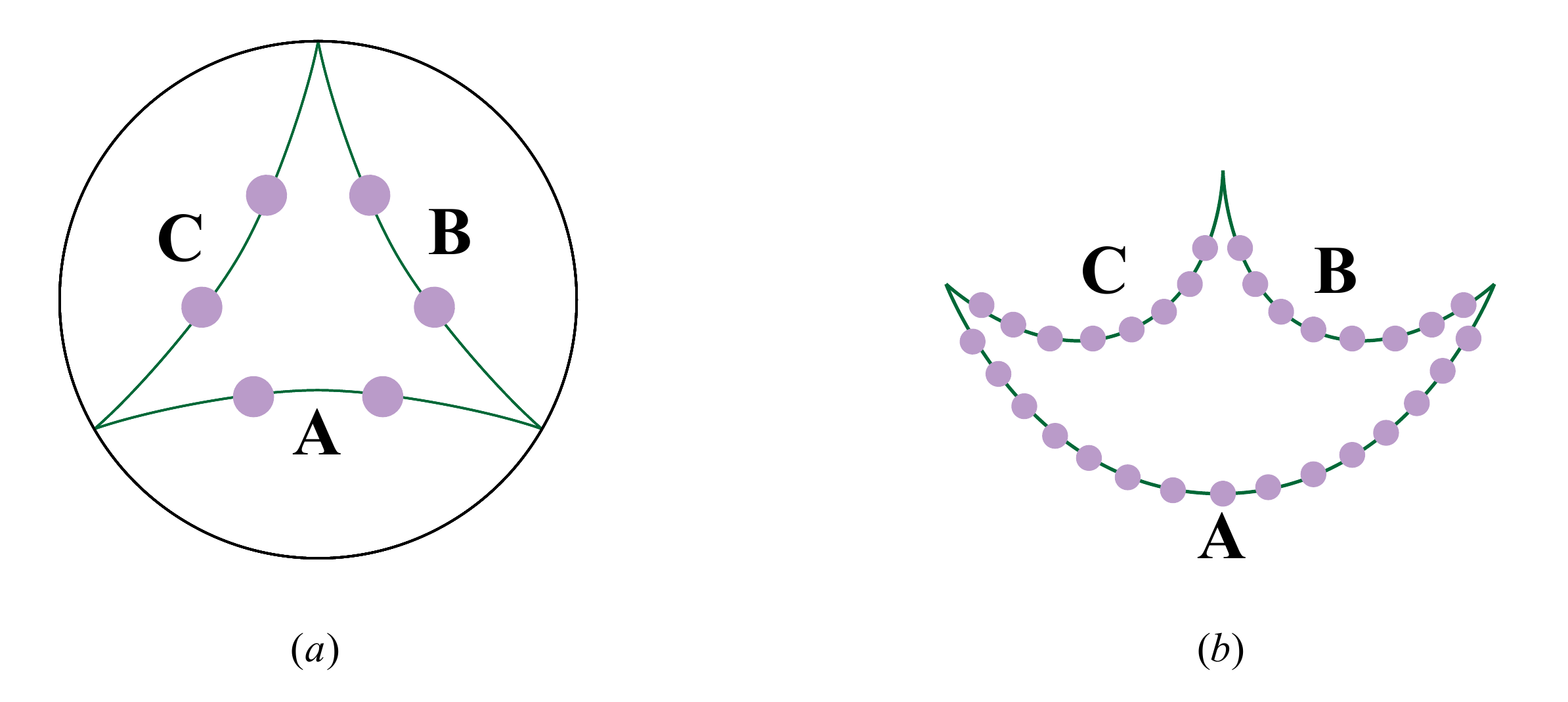}
		\caption{SS bit model: (a) A simple model consisting of three congruent RT surfaces (represented by green lines). Each surface has only two ``SS bits " (represented by mauve discs), and each SS bit has only two states, $\left| 1 \right\rangle$ and $\left| 0 \right\rangle$. (b) A more general case involving ${S_A}$, ${S_B}$, ${S_C}$ SS bits.
		}
		\label{fig4}
	\end{center}	
\end{figure}
The motivation of thread/state correspondence originated from finding an effective scheme to characterize the entanglement structure between different bulk extremal surfaces in the framework of SS duality, or explicitly constructing the dual states (hereinafter referred to as SS states) of these extremal surfaces. This is inspired by the following two notable simple rules in the SS duality~\cite{Miyaji:2015fia,Miyaji:2015yva,Lin:2020ufd}:

$\bf rule ~1:$ the density matrix corresponding to an extremal surface is a direct product of the density matrices at each point. In other words, a bulk extremal surface corresponds to an equal-probability mixed state; 

$\bf rule ~2:$ a closed surface corresponds to a pure state. 

Due to rule 1, one can imagine that on an extremal surface associated with an entropy $S$ (i.e., with an area of $4{G_N}S$), there distribute exactly $S$ bits (each with only two basic states, $\left| 1 \right\rangle $ and $\left| 0 \right\rangle $). Then, for any bulk extremal surface $\gamma$ with an entropy ${S_\gamma }$, we can label its basic states as $\left\{ {\left| {\gamma  = q} \right\rangle \left| {q \in N,0 \le q \le {2^{{S_\gamma }}} - 1} \right.} \right\}$. Each decimal number $q$ exactly corresponds to a binary string that describes the overall configuration state of all the SS bits on $\gamma$. The SS state corresponding to $\gamma$ can be represented as a mixed state ${\rho_\gamma }$ equipped with equal probabilities~\cite{Lin:2020ufd}:
\be\label{label}{\rho _\gamma } = \sum\limits_{q = 0}^{{2^{{S_\gamma }}} - 1} {\frac{1}{{{2^{{S_\gamma }}}}}\left| {\gamma = q} \right\rangle \left\langle {\gamma = q} \right|} .\ee

To illustrate this, as shown in figure~\ref{fig4}(a), les us consider a simple and enlightening model, which is consisting of three congruent RT surfaces. Each surface has only two physical degrees of freedom (let us call them ``SS bits " hereafter), and each SS bit has only two states, i.e., up and down, described by two basic states, $\left| 1 \right\rangle$ and $\left| 0 \right\rangle$, respectively. Let us try to construct the SS states of these three RT surfaces satisfying the two requirements above in the framework of surface/state correspondence. In the present situation, it is required that the corresponding state of the whole system (six SS bits in total) must be a pure state, denoted as $\left| \Psi  \right\rangle $. Moreover, each SS state corresponding to the RT surface must be a direct product mixed state, and in particular, by symmetry, it must be
\be{\rho _{{\rm{ext}}}}{\rm{ = }}\left( {\frac{1}{2}\left| 0 \right\rangle \left\langle 0 \right|{\rm{ + }}\frac{1}{2}\left| 1 \right\rangle \left\langle 1 \right|} \right) \otimes \left( {\frac{1}{2}\left| 0 \right\rangle \left\langle 0 \right|{\rm{ + }}\frac{1}{2}\left| 1 \right\rangle \left\langle 1 \right|} \right),\ee
or
\be{\rho _{{\rm{ext}}}}{\rm{ = }}\frac{1}{4}\left| {00} \right\rangle \left\langle {00} \right|{\rm{ + }}\frac{1}{4}\left| {01} \right\rangle \left\langle {01} \right|{\rm{ + }}\frac{1}{4}\left| {10} \right\rangle \left\langle {10} \right|{\rm{ + }}\frac{1}{4}\left| {11} \right\rangle \left\langle {11} \right|.\ee
Writing it in terms of the mixed state of an ensemble, one can understand what we have described above that each extremal surface corresponds to a mixed state equipped with equal probabilities, i.e., a microcanonical ensemble:
\be{\psi _{{\rm{ext}}}}{\rm{ = }}\left\{ {\left( {\left| {00} \right\rangle ,{p_0} = \frac{1}{4}} \right);\left( {\left| {01} \right\rangle ,{p_1} = \frac{1}{4}} \right);\left( {\left| {10} \right\rangle ,{p_2} = \frac{1}{4}} \right);\left( {\left| {11} \right\rangle ,{p_3} = \frac{1}{4}} \right)} \right\},\ee
The situation is just like that we are counting the possible overall configurations of two non-interacting coins.

Indeed, it is easy to come up with an expression for $\left| \Psi  \right\rangle $ that satisfies the above requirements, that is simply
\be\left| \Psi  \right\rangle {\rm{ = }}\frac{1}{4}\left| {000000} \right\rangle {\rm{ + }}\frac{1}{4}\left| {010101} \right\rangle {\rm{ + }}\frac{1}{4}\left| {101010} \right\rangle {\rm{ + }}\frac{1}{4}\left| {111111} \right\rangle. \ee

Next we move to the more general cases. We will continue with the above simple and natural convention: the physical degrees of freedom on an extremal surface that characterize its corresponding SS state have only two basic states, $\left| 1 \right\rangle $ and $\left| 0 \right\rangle $. This is similar to the case in holographic tensor networks, where the state corresponding to an RT surface is defined as the direct product state of $n$ bonds cut off by the surface. Assigning each bond a Hilbert space dimension $J$, the Von Neumann entropy corresponding to the extremal surface is given by~\cite{Swingle:2009bg,Swingle:2012wq}
\be S = n \cdot {\ln _2}J.\ee
Our SS bit model is equivalent to choosing $J=2$, such that
\be S = S \cdot {\ln _2}2,\ee
that is, for an extremal surface associated with an entropy $S$ (i.e., with an area of $4{G_N}S$), the number of SS bits needed in our model is exactly $S$.

As shown in figure~\ref{fig4}(b), we still consider the case consisting of three RT surfaces, denoted as $A$, $B$, and $C$ respectively. However, now there are ${S_A}$, ${S_B}$, ${S_C}$ SS bits associated with these extremal surfaces respectively, which correspond to the areas of the three surfaces being $4{G_N}{S_A}$, $4{G_N}{S_B}$ and $4{G_N}{S_C}$ respectively. Naturally, each basic state of the RT surfaces can be denoted by a binary number. For example, surface $A$ can be in state $\left| {\underbrace {00 \cdots 00}_{{S_A}}} \right\rangle  \equiv \left| {A = 0} \right\rangle$, $\left| {\underbrace {00 \cdots 01}_{{S_A}}} \right\rangle  \equiv \left| {A = 1} \right\rangle$, $\left| {\underbrace {00 \cdots 10}_{{S_A}}} \right\rangle  \equiv \left| {A = 2} \right\rangle $, and so on, all the way to $\left| {\underbrace {11 \cdots 11}_{{S_A}}} \right\rangle  \equiv \left| {A = {2^{{S_A}}} - 1} \right\rangle$. This allows us to label each basic state of surface $A$ in a convenient way, that is, $\left\{ {\left| {A = q} \right\rangle \left| {q \in N,0 \le q \le {2^{{S_A}}} - 1} \right.} \right\}$. Each decimal number $q$ corresponds to a binary string that describes the overall configuration state of all the SS bits on $A$. More generally, for any bulk extremal surface $\gamma$ with an entropy ${S_\gamma }$, we thus can label its basic states as the form of~(\ref{label}), where each state $\left| {\gamma  = q} \right\rangle$ can be called a ``configuration eigenstate'' properly, because the number $q$ can be converted to binary to indicate the configuration state (1 or 0) of each individual SS bit on the surface $\gamma$. Number $q$ also happens to mean that the surface is in the $q$th configuration state, going from the 0th to the ${2^{{S_\gamma }}} – 1$th, so there are ${2^{{S_\gamma }}}$ possible configurations in total.

So, how to find a pure state $\left| \Psi  \right\rangle$ (rather than a mixed state) corresponding to the closed surface consisting of $A$, $B$, and $C$ in figure~\ref{fig4}(b)? This is a typical entanglement of purification problem. A natural way is to adopt the idea of Schmidt's decomposition. From the full pure state $\left| \Psi  \right\rangle$, we are going to read the correct mixed state ${\psi _A}$, which can be written in the form of a reduced density matrix as
\be\label{rhoa}{\rho _A} = \sum\limits_{{q_a} = 0}^{{2^{{S_A}}} - 1} {\frac{1}{{{2^{{S_A}}}}}\left| {A = {q_a}} \right\rangle \left\langle {A = {q_a}} \right|} .\ee
Therefore, we can construct 
\be\label{pure}\left| \Psi  \right\rangle  = \sum\limits_{q = 0}^{{2^{{S_A}}} - 1} {\frac{1}{{\sqrt {{2^{{S_A}}}} }}\left| {A = q} \right\rangle  \otimes \left| {BC = q} \right\rangle } ,\ee
where $\left| {BC = q} \right\rangle $ is merely a formal definition at present, representing the basic states of the complement of region $A$ in the whole closed system.

Here is an illuminating attempt. Naively, one might think that $\left| {BC = q} \right\rangle$ could be simply chosen to be some certain overall configuration state of the composite system of $B$ and $C$, such as $\left| {BC = 0} \right\rangle  = \left| {\underbrace {00 \cdots 00}_{{S_B}}\underbrace {00 \cdots 00}_{{S_C}}} \right\rangle  \equiv \left| {\underbrace {00 \cdots 00}_{{S_B}}} \right\rangle  \otimes \left| {\underbrace {00 \cdots 00}_{{S_C}}} \right\rangle $. However, from the full pure state φ, we should not only be able to work out ${\psi _A}$ as a mixed state consisting of ${2^{{S_A}}}$ equal-probability configurations, but also work out  ${\psi _B}$ as a mixed state consisting of ${2^{{S_B}}}$ equal-probability configurations, and similarly, ${\psi _C}$ as a mixed state consisting of ${2^{{S_C}}}$ equal-probability configurations. In order for $\left| \Psi  \right\rangle$ to accommodate so much information of configurations, we realize that each $\left| {BC = q} \right\rangle$ state should actually be a quantum superposition state. A natural choice is to construct $\left| {BC = q} \right\rangle$ as a quantum superposition of ${2^{{S_B} + {S_C} - {S_A}}}$ configuration eigenstates of the  composite system of $B$ and $C$. In this way, since we have ${2^{{S_A}}}$ $\left| {A = q} \right\rangle$ states, it is possible for the full pure state $\left| \Psi  \right\rangle$  to contain the information of a total of ${2^{{S_B} + {S_C}}}$ possible overall configurations of $BC$.  Furthermore, based on symmetry (more specifically, on the equality of every SS bit), each amplitude in the superposition would be expected to be equal. Therefore, formally, we can construct the following ansatz:
\begin{tiny}\be\label{ans}\left| {BC = q} \right\rangle  = \frac{1}{{\sqrt {{2^{{S_B} + {S_C} - {S_A}}}} }}\left( {\underbrace {\left| {\underbrace { -  -  \cdots  -  - }_{{S_B}}\underbrace { -  -  \cdots  -  - }_{{S_C}}} \right\rangle  + \left| {\underbrace { -  -  \cdots  -  - }_{{S_B}}\underbrace { -  -  \cdots  -  - }_{{S_C}}} \right\rangle  +  \cdots  + \left| {\underbrace { -  -  \cdots  -  - }_{{S_B}}\underbrace { -  -  \cdots  -  - }_{{S_C}}} \right\rangle }_{{2^{{S_B} + {S_C} - {S_A}}}}} \right),\ee\end{tiny}
where $\left| {\underbrace { -  -  \cdots  -  - }_{{S_B}}\underbrace { -  -  \cdots  -  - }_{{S_C}}} \right\rangle$ represents some certain configuration eigenstate of composite system $BC$. Thus, we can more explicitly write the normalized pure state of the whole system as
\begin{tiny}
\be\label{full}\begin{array}{l}
	\left| \Psi  \right\rangle  = \frac{1}{{\sqrt {{2^{{S_A}}}} }}\left| {A = 0} \right\rangle  \otimes \left\{ {\frac{1}{{\sqrt {{2^{{S_B} + {S_C} - {S_A}}}} }}\left( {\underbrace {\left| {\underbrace { -  -  \cdots  -  - }_{{S_B}}\underbrace { -  -  \cdots  -  - }_{{S_C}}} \right\rangle  + \left| {\underbrace { -  -  \cdots  -  - }_{{S_B}}\underbrace { -  -  \cdots  -  - }_{{S_C}}} \right\rangle  +  \cdots  + \left| {\underbrace { -  -  \cdots  -  - }_{{S_B}}\underbrace { -  -  \cdots  -  - }_{{S_C}}} \right\rangle }_{{2^{{S_B} + {S_C} - {S_A}}}}} \right)} \right\}\\
	+ \frac{1}{{\sqrt {{2^{{S_A}}}} }}\left| {A = 1} \right\rangle  \otimes \left\{ {\frac{1}{{\sqrt {{2^{{S_B} + {S_C} - {S_A}}}} }}\left( {\underbrace {\left| {\underbrace { -  -  \cdots  -  - }_{{S_B}}\underbrace { -  -  \cdots  -  - }_{{S_C}}} \right\rangle  + \left| {\underbrace { -  -  \cdots  -  - }_{{S_B}}\underbrace { -  -  \cdots  -  - }_{{S_C}}} \right\rangle  +  \cdots  + \left| {\underbrace { -  -  \cdots  -  - }_{{S_B}}\underbrace { -  -  \cdots  -  - }_{{S_C}}} \right\rangle }_{{2^{{S_B} + {S_C} - {S_A}}}}} \right)} \right\}\\
	+  \cdots \\
	+ \frac{1}{{\sqrt {{2^{{S_A}}}} }}\left| {A = {2^{{S_A}}} - 1} \right\rangle  \otimes \left\{ {\frac{1}{{\sqrt {{2^{{S_B} + {S_C} - {S_A}}}} }}\left( {\underbrace {\left| {\underbrace { -  -  \cdots  -  - }_{{S_B}}\underbrace { -  -  \cdots  -  - }_{{S_C}}} \right\rangle  + \left| {\underbrace { -  -  \cdots  -  - }_{{S_B}}\underbrace { -  -  \cdots  -  - }_{{S_C}}} \right\rangle  +  \cdots  + \left| {\underbrace { -  -  \cdots  -  - }_{{S_B}}\underbrace { -  -  \cdots  -  - }_{{S_C}}} \right\rangle }_{{2^{{S_B} + {S_C} - {S_A}}}}} \right)} \right\}
\end{array},\ee
\end{tiny}
where we denote $\begin{tiny}\left| {\underbrace { -  -  \cdots  -  - }_{{S_B}}\underbrace { -  -  \cdots  -  - }_{{S_C}}} \right\rangle  = \left| {\underbrace { -  -  \cdots  -  - }_{{S_B}}} \right\rangle  \otimes \left| {\underbrace { -  -  \cdots  -  - }_{{S_C}}} \right\rangle  \equiv \left| {B = {q_b}} \right\rangle  \otimes \left| {C = {q_c}} \right\rangle\end{tiny}$. Note that this is indeed a pure state, because it is merely a quantum superposition of a set of basic states, rather than a mixed state. On the other hand, by calculating the reduced density matrixes for each subregion $A$, $B$, and $C$ respectively, or equivalently, reading the probability distributions of the reduced states directly from the above Schmidt form, one can find that each individual RT surface corresponds to a mixed state respectively. In particular, for the surface $A$, it is clear that the expression  (\ref{rhoa}) is satisfied. Actually, we still have not determined the exact form of each state in (\ref{full}). Anyway, we hope to obtain
\be\label{rhob}{\rho _B} = \sum\limits_{{q_b} = 0}^{{2^{{S_B}}} - 1} {\frac{1}{{{2^{{S_B}}}}}\left| {B = {q_b}} \right\rangle \left\langle {B = {q_b}} \right|} ,\ee
and
\be\label{rhoc}{\rho _C} = \sum\limits_{{q_c} = 0}^{{2^{{S_C}}} - 1} {\frac{1}{{{2^{{S_C}}}}}\left| {C = {q_c}} \right\rangle \left\langle {C = {q_c}} \right|} .\ee 
The point is that, by simply putting all possible configuration states $\left| {B = {q_b}} \right\rangle  \otimes \left| {C = {q_c}} \right\rangle $ into the big curly braces of (\ref{full}), then the expression (\ref{full}) is enough to fulfil the task of suitably characterizing the pure state corresponding to the closed surface $ABC$ in some sense. To see this, note that in principle we can use the conditions (\ref{rhob}) (\ref{rhoc}) to determine the information of how the items are distributed in the curly braces of (\ref{full}). For example, we can determine how many of the items in the curly braces of (\ref{full}) should contain $\left| {B = 0} \right\rangle  = \left| {\underbrace {00 \cdots 00}_{{S_B}}} \right\rangle $ as a partial configuration. For instance, more specifically, from (\ref{rhob}), we can determine that the number of $\begin{tiny}\left| {\underbrace {00 \cdots 00}_{{S_B}}\underbrace { -  -  \cdots  -  - }_{{S_C}}} \right\rangle  = \left| {\underbrace {00 \cdots 00}_{{S_B}}} \right\rangle  \otimes \left| {\underbrace { -  -  \cdots  -  - }_{{S_C}}} \right\rangle  \equiv \left| {B = 0} \right\rangle  \otimes \left| {C = {q_c}} \right\rangle\end{tiny} $ that should appear in the curly braces of (\ref{full}) should be 
\be\# \left( {\left| {\underbrace {00 \cdots 00}_{{S_B}}\underbrace { -  -  \cdots  -  - }_{{S_C}}} \right\rangle } \right) = \frac{{\frac{1}{{{2^{{S_B}}}}}}}{{{{\left( {\frac{1}{{\sqrt {{2^{{S_A}}}} }} \times \frac{1}{{\sqrt {{2^{{S_B} + {S_C} - {S_A}}}} }}} \right)}^2}}} = {2^{{S_C}}}\ee
Obviously, since there are only a total of ${2^{{S_C}}}$ different configuration choices for $\left| {C = {q_c}} \right\rangle $, the curly braces of (\ref{full}) should include exactly all these terms, i.e.: $\left| {\underbrace {00 \cdots 00}_{{S_B}}\underbrace {00 \cdots 00}_{{S_C}}} \right\rangle $, $\left| {\underbrace {00 \cdots 00}_{{S_B}}\underbrace {00 \cdots 01}_{{S_C}}} \right\rangle $, $\left| {\underbrace {00 \cdots 00}_{{S_B}}\underbrace {00 \cdots 10}_{{S_C}}} \right\rangle $, $\cdots$, $\left| {\underbrace {00 \cdots 00}_{{S_B}}\underbrace {11 \cdots 11}_{{S_C}}} \right\rangle$. Using the same analysis, we realized that putting all possible $\left| {B = {q_b}} \right\rangle  \otimes \left| {C = {q_c}} \right\rangle$ configuration eigenstates into the curly braces of (\ref{full}) will nicely complete the assignment. Indeed, the number of undetermined items in the curly braces exactly matches the number of possible configurations for $\left| {B = {q_b}} \right\rangle  \otimes \left| {C = {q_c}} \right\rangle$, i.e., ${2^{{S_B} + {S_C}}}$.

%%%%%%%%%%%%%%%%%%%%%%%%%%%%%%%%%%%%%%%%%%%%%%%%%%%%%%%%%%%%%%%%%%%%%%

\subsection{Prescription: thread/state correspondence}\label{subsec3.2}
%%%%%%%%%%%%%%%%%%%%%%%%%%%%%%%%%%%%%%%%%%%%%%%%%%%%%%%%%%%%%%%%%%%%%%
\begin{figure}[htbp]     \begin{center}
		\includegraphics[height=10cm,clip]{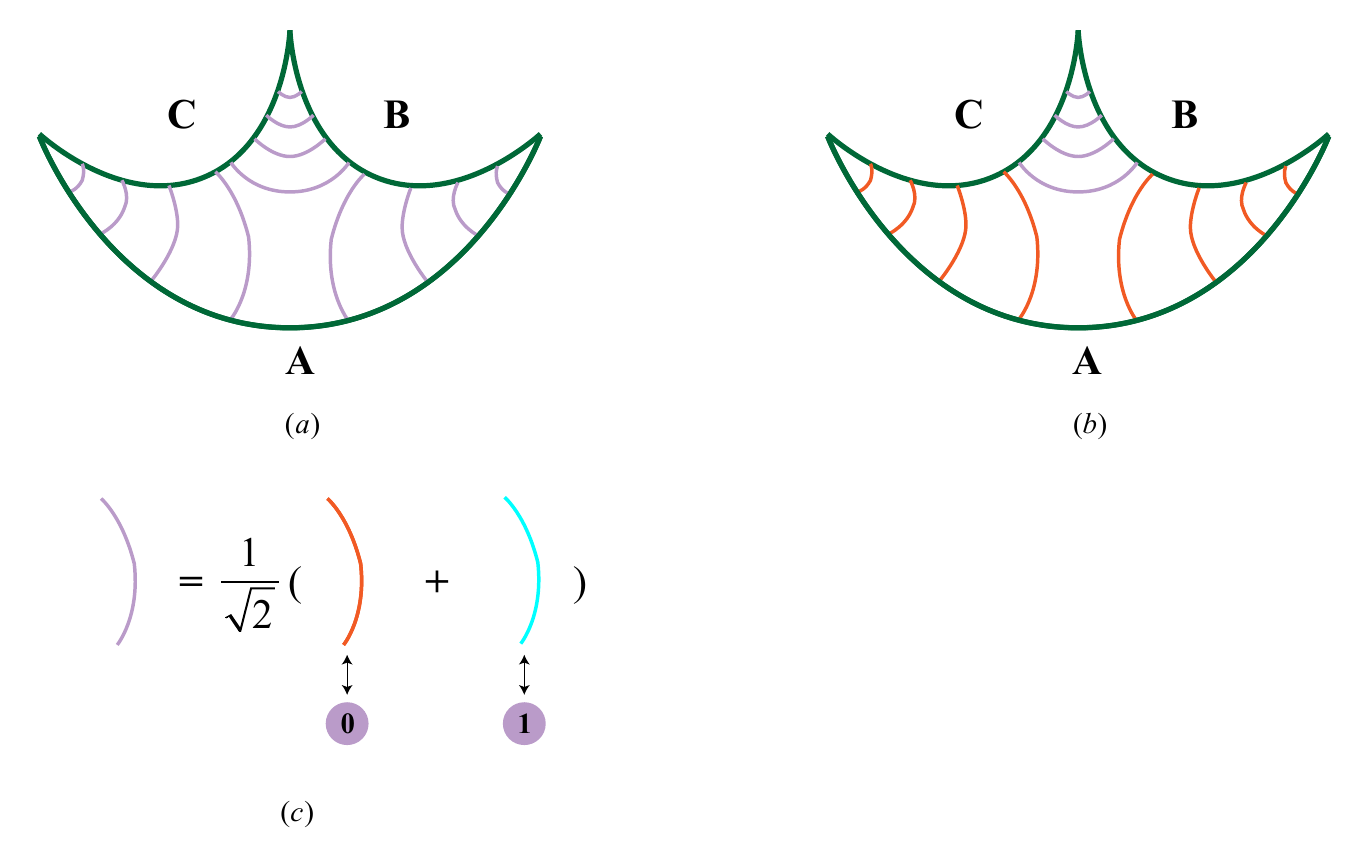}
		\caption{(a) A locking bit thread configuration can automatically generate the SS states of a set of RT surfaces (green lines). (b) If we measure surface $A$ to be in the state $\left| {A = 0} \right\rangle  $, then all bit threads passing through surface $A$ are accordingly in red state. (c) Thread/state correspondence. Each bit thread has two possible orthogonal states, represented by red lines and blue lines respectively. While the mauve lines represent the bit threads in the quantum superposition state.}
		\label{fig5}
	\end{center}	
\end{figure}

The question remains, however, of how to fit all the possible $\left| {B = {q_b}} \right\rangle  \otimes \left| {C = {q_c}} \right\rangle$ configuration states into that big curly brace in a natural way. One of our interesting discoveries is that the so-called locking bit thread configuration~\cite{Headrick:2020gyq} can automatically implement this task, albeit in a slightly different way. Moreover, our discovery will endow bit threads with a definite physical meaning. Later, we will also explore an interesting picture about spacetime emergence. Consider a locking bit thread configuration that can lock three regions $A$, $B$, and $C$ (which are three RT surfaces) simultaneously, as shown in figure~\ref{fig5}(a). It is characterized by a multiflow consisting of three component flows ${v_{AB}}$, ${v_{AC}}$ and ${v_{BC}}$, each of which describes an independent thread bundle. Define  
\be F{(\alpha )_{ij}} = \left| {\int_\alpha  {{{\vec v}_{ij}}} } \right| = \left| {\int_\alpha  {\sqrt h {{\hat n}_\alpha } \cdot {{\vec v}_{ij}}} } \right|\ee
to represent the value of the flux of the bit threads described by the component flow ${\vec v_{ij}}$ passing through the $\alpha $ surface, where $h$ is the determinant of the induced metric on the surface $\alpha $, and ${\hat n_\alpha }$ is the unit normal vector on surface $\alpha $. Due to the divergenceless property of bit threads, the $F{(\alpha )_{ij}}$ associated with each component flow does not depend on the surfaces the threads pass through in the bulk, and thus can be abbreviated as $ {F_{ij}} $. In the framework of the locking bit thread scheme, the flux $ {F_{ij}} $ of each thread bundle in a locking thread configuration characterizes the amount of the entanglement between the two regions it connects. In this simple case, it is easy to obtain~\cite{Lin:2020yzf}
\be\label{inf}\begin{array}{l}
	{F_{AB}} = \frac{1}{2}\left( {{S_A} + {S_B} - {S_C}} \right) \equiv \frac{1}{2}{I_{AB}}\\
	{F_{AC}} = \frac{1}{2}\left( {{S_A} + {S_C} - {S_B}} \right) \equiv \frac{1}{2}{I_{AC}}\\
	{F_{BC}} = \frac{1}{2}\left( {{S_B} + {S_C} - {S_A}} \right) \equiv \frac{1}{2}{I_{BC}}
\end{array},\ee
where $I$ is mutual information. 

We are now going to introduce a property of bit threads that has not been explicitly mentioned before. We propose that each bit thread can be in two orthogonal states, which we vividly name as red state and blue state. More specifically, we propose the following thread /state correspondence: 

$(\bf 1)$ Each bit thread has two possible orthogonal states, namely $\left| {{\rm{red}}} \right\rangle $ state and $\left| {{\rm{blue}}} \right\rangle$ state. For convenience, we will also directly refer to the bit threads in the two states as red bit threads and blue bit threads intuitively. The same bit thread is always in a same ``color'' state, or a same superposition state of the color states, while different bit threads are independent of each other, that is, the state of one bit thread does not affect the state of the other bit thread.

$(\bf 2)$ In a locking bit thread configuration, each bit thread is in a special quantum superposition state (for convenience, we use the mauve lines to represent the bit threads in this superposition in the figure~\ref{fig5}):
\be\label{thr}\left| {{\rm{thread}}} \right\rangle  = \frac{1}{{\sqrt 2 }}\left( {\left| {{\rm{red}}} \right\rangle  + \left| {{\rm{blue}}} \right\rangle } \right)\ee

$(\bf 3)$ In a locking bit thread configuration, the bit threads exactly meet the SS bits on the intersecting bulk extremal surfaces. The red state of a bit thread corresponds to the $\left| 0 \right\rangle$ state of the SS bits it passes through on the bulk extremal surfaces, while the blue state corresponds to the $\left| 1 \right\rangle $ state. In fact, this correspondence can be written more explicitly as
\be\begin{array}{l}
	\left| {{\rm{red}}} \right\rangle  = \left| {00 \cdots 00} \right\rangle \\
	\left| {{\rm{blue}}} \right\rangle  = \left| {11 \cdots 11} \right\rangle 
\end{array}\ee
In other words, if we perform a measurement operation to make a bit thread in a certain color state, the SS bits crossed by the same red bit thread will be all in the certain $\left| 0 \right\rangle$ state, while the SS bits crossed by the same blue bit thread will be all in the certain $\left| 1 \right\rangle$ state. 

According to this, (\ref{thr}) can also be written as
\be\label{thr2}\left| {{\rm{thread}}} \right\rangle  = \frac{1}{{\sqrt 2 }}\left( {\left| {00 \cdots 00} \right\rangle  + \left| {11 \cdots 11} \right\rangle } \right).\ee
It follows that, as expected, each SS bit passed through by bit threads can be exactly characterized by a reduced density matrix 
\be{\rho _{{\rm{bit}}}}{\rm{ = }}\frac{1}{2}\left| 0 \right\rangle \left\langle 0 \right| + \frac{1}{2}\left| 1 \right\rangle \left\langle 1 \right|.\ee
It turns out that, with such a prescription of thread/state correspondence, the locking bit thread configurations can be dexterously and naturally used to characterize the entanglement structure between the SS states of a set of bulk extremal surface in the surface/state correspondence.

Now we will show that, using these thread/state rules, we will end up with an explicit form of the state of the whole closed surface $ABC$ as:
\be\label{fullr} \begin{array}{l}
	\left| \Psi  \right\rangle  = \sum\limits_{{q_{AB}} = 0}^{{2^{{F_{AB}}}} - 1} {\sum\limits_{{q_{AC}} = 0}^{{2^{{F_{AC}}}} - 1} {\sum\limits_{{q_{BC}} = 0}^{{2^{{F_{BC}}}} - 1} {\frac{1}{{\sqrt {{2^{{F_{AB}} + {F_{AC}} + {F_{BC}}}}} }}} } } \\
	\cdot \left( {\left| {{\phi _{AB}} = {q_{AB}}} \right\rangle  \otimes \left| {{\phi _{AC}} = {q_{AC}}} \right\rangle } \right)\\
	\otimes \left( {\left| {{\phi _{AB}} = {q_{AB}}} \right\rangle  \otimes \left| {{\phi _{BC}} = {q_{BC}}} \right\rangle } \right)\\
	\otimes \left( {\left| {{\phi _{AC}} = {q_{AC}}} \right\rangle  \otimes \left| {{\phi _{BC}} = {q_{BC}}} \right\rangle } \right)
\end{array}, \ee
where $\left| {{\phi _{AB}} = {q_{AB}}} \right\rangle  = \left| {\overbrace { -  -  \cdots  -  - }^{{F_{AB}}}} \right\rangle $ represents a configuration state of the bit threads connecting $A$ and $B$, whose number is ${F_{AB}}$, and it also provides the information of the configuration states of the SS bits on the bulk extremal surfaces that these bit threads pass through. The similar notation is used for ${\phi _{AC}}$ and ${\phi _{BC}}$. Note that (\ref{fullr}) is indeed a pure state, and the key point is that it is not difficult to verify that this expression (\ref{fullr}) can indeed be reduced to the correct reduced density matrixes (\ref{label}) of $A$, $B$, and $C$.

Before we dive deeper into details, highlighting two important points in more physical language will help to understand why our prescription works. Let us firstly focus on a single subregion, such as RT surface $A$, we say, the von Neumann entropy of $A$ measures the amount of quantum entanglement between $A$ and the complement $BC$. When we say that there is quantum entanglement between $A$ and $BC$, in more physical language, it means that when we {\bf measure} that $A$ is in some certain configuration $\left| {A = i} \right\rangle$, then accordingly $BC$ must be in some corresponding configuration $\left| {BC = i} \right\rangle$.~\footnote{ This is in contrast to the case of two coins ``without quantum entanglement'' in the everyday classical world, where whether we {\bf measure} one of the coins to be up or down does not {\bf determine} whether the other coin must be up or down.} Secondly, we say, since $ABC$ is in a pure state, the Von Neumann entropy of $BC$ as a whole is exactly equal to the von Neumann entropy of $A$. Physically, this implies that the number of possible configuration choices for $BC$ as a whole is exactly the same as the number of possible configuration choices for $A$. However, at first glance, the number of possible configurations of $B$ times the number of possible configurations of $C$ is obviously greater than the number of possible configurations of $A$. This is also, of course, because there is also the constraint of quantum entanglement between $B$ and $C$, so that when a part of SS bits on $B$ is in some certain states, a part of SS bits on $C$ must also accordingly be in some corresponding states.

Now we present the details (a user guide to thread/state prescription). Focusing on the correct equal-probability mixed states~(\ref{rhoa}), the pure state corresponding to the whole closed surface $ABC$ can be constructed as~(\ref{pure}). As shown in figure~\ref{fig5}(b), suppose we are going to measure surface $A$ and find that $A$ is in a certain configuration after the measurement operation, for example, in the state $\left| {A = 0} \right\rangle  \equiv \left| {\underbrace {00 \cdots 00}_{{S_A}}} \right\rangle $. Then by {\bf rule 3}, this is also equivalent to measuring that all bit threads passing through surface $A$ (i.e., starting from $A$, and connecting to $B$ or $C$) are in red state. As per {\bf rule 1}, the operation of measuring $A$ will not affect the state of other bit threads that do not cross $A$ (i.e. threads connecting $B$ with $C$), then we can immediately construct the explicit form of the corresponding configuration$\left| {BC = 0} \right\rangle$ that $BC$ must be in when $A$ is measured to be  in configuration $\left| {A = 0} \right\rangle  \equiv \left| {\underbrace {00 \cdots 00}_{{S_A}}} \right\rangle $. Firstly, under this measurement, that $A$ is in configuration $\left| {A = 0} \right\rangle  \equiv \left| {\underbrace {00 \cdots 00}_{{S_A}}} \right\rangle$ corresponds to the bit threads passing through surface $A$ (to $B$ or $C$) are all in red state, therefore, in fact, we have determined the states that a certain part of the SS bits on surface $B$ and $C$ must be in, that is, the SS bits passed through by these red bit threads are all in the definite state $\left| 0 \right\rangle $. On the other hand, the other bits on $B$ and $C$ are passed through by the bit threads in the superposition state (\ref{thr}) ({\bf rule 2}). Therefore, the prescription automatically produces the specific form of $\left| {BC = 0} \right\rangle $ as
\be\begin{array}{l}
	\left| {BC = 0} \right\rangle  = \chi  \cdot \{ \left| {\underbrace {\overbrace {00 \cdots 00}^{{F_{AB}}}\overbrace {00 \cdots 00}^{{F_{BC}}}}_{{S_B}}\underbrace {\overbrace {00 \cdots 00}^{{F_{AC}}}\overbrace {00 \cdots 00}^{{F_{BC}}}}_{{S_C}}} \right\rangle \\
	+ \left| {\underbrace {\overbrace {00 \cdots 00}^{{F_{AB}}}\overbrace {00 \cdots 01}^{{F_{BC}}}}_{{S_B}}\underbrace {\overbrace {00 \cdots 00}^{{F_{AC}}}\overbrace {00 \cdots 01}^{{F_{BC}}}}_{{S_C}}} \right\rangle \\
	+  \cdots  + \left| {\underbrace {\overbrace {00 \cdots 00}^{{F_{AB}}}\overbrace {11 \cdots 11}^{{F_{BC}}}}_{{S_B}}\underbrace {\overbrace {00 \cdots 00}^{{F_{AC}}}\overbrace {11 \cdots 11}^{{F_{BC}}}}_{{S_C}}} \right\rangle \} 
\end{array},\ee

where $\chi $ is a normalization constant that should satisfy
\be{\chi ^2} \cdot {2^{{F_{BC}}}} = 1,\ee
that is,
\be\chi  = \frac{1}{{\sqrt {{2^{{F_{BC}}}}} }} = \frac{1}{{\sqrt {{2^{\frac{{{S_B} + {S_C} - {S_A}}}{2}}}} }}.\ee

Similarly, we can use the same reasoning to obtain the explicit form of the corresponding $\left| {BC = q} \right\rangle $ state for each $\left| {A = q} \right\rangle $ state, and finally obtain the expression of the full pure state of the whole system $ABC$ as follows:
\be\label{full2}\begin{array}{l}
		\left| \Psi  \right\rangle  =\\
		\frac{1}{{\sqrt {{2^{{F_{AB}} + {F_{AC}}}}} }}\left| {\underbrace {\overbrace {00 \cdots 00}^{{F_{AB}}}\overbrace {00 \cdots 00}^{{F_{AC}}}}_{{S_A}}} \right\rangle  \otimes \{ \frac{1}{{\sqrt {{2^{{F_{BC}}}}} }}(\left| {\underbrace {\overbrace {00 \cdots 00}^{{F_{AB}}}\overbrace {00 \cdots 00}^{{F_{BC}}}}_{{S_B}}\underbrace {\overbrace {00 \cdots 00}^{{F_{AC}}}\overbrace {00 \cdots 00}^{{F_{BC}}}}_{{S_C}}} \right\rangle \\
		+ \left| {\underbrace {\overbrace {00 \cdots 00}^{{F_{AB}}}\overbrace {00 \cdots 01}^{{F_{BC}}}}_{{S_B}}\underbrace {\overbrace {00 \cdots 00}^{{F_{AC}}}\overbrace {00 \cdots 01}^{{F_{BC}}}}_{{S_C}}} \right\rangle  +  \cdots  + \left| {\underbrace {\overbrace {00 \cdots 00}^{{F_{AB}}}\overbrace {11 \cdots 11}^{{F_{BC}}}}_{{S_B}}\underbrace {\overbrace {00 \cdots 00}^{{F_{AC}}}\overbrace {11 \cdots 11}^{{F_{BC}}}}_{{S_C}}} \right\rangle )\} \\
		+ \frac{1}{{\sqrt {{2^{{F_{AB}} + {F_{AC}}}}} }}\left| {\underbrace {\overbrace {00 \cdots 00}^{{F_{AB}}}\overbrace {00 \cdots 01}^{{F_{AC}}}}_{{S_A}}} \right\rangle  \otimes \{ \frac{1}{{\sqrt {{2^{{F_{BC}}}}} }}(\left| {\underbrace {\overbrace {00 \cdots 00}^{{F_{AB}}}\overbrace {00 \cdots 00}^{{F_{BC}}}}_{{S_B}}\underbrace {\overbrace {00 \cdots 01}^{{F_{AC}}}\overbrace {00 \cdots 00}^{{F_{BC}}}}_{{S_C}}} \right\rangle \\
		+ \left| {\underbrace {\overbrace {00 \cdots 00}^{{F_{AB}}}\overbrace {00 \cdots 01}^{{F_{BC}}}}_{{S_B}}\underbrace {\overbrace {00 \cdots 01}^{{F_{AC}}}\overbrace {00 \cdots 01}^{{F_{BC}}}}_{{S_C}}} \right\rangle  +  \cdots  + \left| {\underbrace {\overbrace {00 \cdots 00}^{{F_{AB}}}\overbrace {11 \cdots 11}^{{F_{BC}}}}_{{S_B}}\underbrace {\overbrace {00 \cdots 01}^{{F_{AC}}}\overbrace {11 \cdots 11}^{{F_{BC}}}}_{{S_C}}} \right\rangle )\} \\
		+  \cdots \\
		\frac{1}{{\sqrt {{2^{{F_{AB}} + {F_{AC}}}}} }}\left| {\underbrace {\overbrace {11 \cdots 11}^{{F_{AB}}}\overbrace {11 \cdots 11}^{{F_{AC}}}}_{{S_A}}} \right\rangle  \otimes \{ \frac{1}{{\sqrt {{2^{{F_{BC}}}}} }}(\left| {\underbrace {\overbrace {11 \cdots 11}^{{F_{AB}}}\overbrace {00 \cdots 00}^{{F_{BC}}}}_{{S_B}}\underbrace {\overbrace {11 \cdots 11}^{{F_{AC}}}\overbrace {00 \cdots 00}^{{F_{BC}}}}_{{S_C}}} \right\rangle \\
		+ \left| {\underbrace {\overbrace {11 \cdots 11}^{{F_{AB}}}\overbrace {00 \cdots 01}^{{F_{BC}}}}_{{S_B}}\underbrace {\overbrace {11 \cdots 11}^{{F_{AC}}}\overbrace {00 \cdots 01}^{{F_{BC}}}}_{{S_C}}} \right\rangle  +  \cdots  + \left| {\underbrace {\overbrace {11 \cdots 11}^{{F_{AB}}}\overbrace {11 \cdots 11}^{{F_{BC}}}}_{{S_B}}\underbrace {\overbrace {11 \cdots 11}^{{F_{AC}}}\overbrace {11 \cdots 11}^{{F_{BC}}}}_{{S_C}}} \right\rangle )\} 
	\end{array}.\ee
Notice that therein ${F_{AB}} + {F_{AC}} = {S_A}$. Rewriting (\ref{full2}) compactly then results  in the simple and symmetric form (\ref{fullr}).

This expression is not exactly the same as the expression (\ref{full}). However, it is easy to check that the expression (\ref{full2}) can also meet the requirements of purifying (\ref{rhoa}), (\ref{rhob}) and (\ref{rhoc}). The most significant difference between the two purification schemes (\ref{full}) and (\ref{full2}) is that the former involves a total of ${2^{{S_B} + {S_C}}}$ basic states, while the latter has a total of ${2^{{F_{AB}} + {F_{AC}} + {F_{BC}}}}$ basic states. Although from a mathematical point of view, the previous scheme can also implement the task, from a physical point of view, the present scheme is more symmetrical, wherein the ${F_{AB}}$, ${F_{AC}}$ and ${F_{BC}}$  are playing equal roles. Furthermore, it is more satisfactory to see that the present scheme has a simple and beautiful physical picture dipicted by bit threads. 

A physical comment: back to the second point we emphasized at the beginning of this section, namely, because of the quantum entanglement between $B$ and $C$, the Hilbert space dimension of the states of $BC$ as a whole is exactly equal to the Hilbert space dimension of the corresponding states of $A$. Our thread /state correspondence rules explicitly show how the bit threads characterize the quantum entanglement between $B$ and $C$: that is, the SS bits on surfaces $B$ and $C$ crossed by the same bit thread must be in the same state. From another point of view, our thread/state rules explicitly accentuate for the first time the implied meaning of the name ``bit threads'', namely, that these threads, which can be used mathematically to recover the RT formula, can be physically assigned a meaning closely related to the concept of ``bits''.

To conclude, under the framework of surface/state duality, a locking bit thread configurations is an effective prescription that can automatically generates the SS states (which are the states in the Hilbert space of the holographic boundary CFT) dual to a set of specified bulk extremal surfaces and furthermore, manifests the entanglement relations between these SS states. We can also formulate this picture in another way: In the language of ``It from qubit", the information of bulk spacetime is encoded in the information of the boundary CFT state, then a locking bit thread configuration can be regarded as an effective decoding approach. In this way, we can simultaneously read and extract the information of a special set of extremal surfaces in the holographic bulk spacetime and their entanglement relations with each other.

%%%%%%%%%%%%%%%%%%%%%%%%%%%%%%%%%%%%%%%%%%%%%%%%%%%%%%%%%%%%%%%%%%%%%%
\section{holographic qubit threads model}\label{sec4}
%%%%%%%%%%%%%%%%%%%%%%%%%%%%%%%%%%%%%%%%%%%%%%%%%%%%%%%%%%%%%%%%%%%%%%

%%%%%%%%%%%%%%%%%%%%%%%%%%%%%%%%%%%%%%%%%%%%%%%%%%%%%%%%%%%%%%%%%%%%%%
\subsection{Reading SS states from holographic qubit threads model }\label{subsec4.1}
%%%%%%%%%%%%%%%%%%%%%%%%%%%%%%%%%%%%%%%%%%%%%%%%%%%%%%%%%%%%%%%%%%%%%%
\begin{figure}[htbp]     \begin{center}
		\includegraphics[height=10cm,clip]{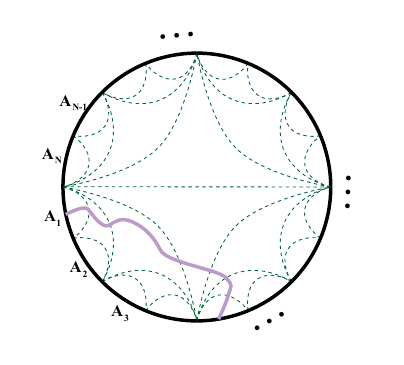}
		\caption{The holographic qubit threads model. The boundary quantum system is divided into extremely large $N$ elementary regions (of the same size), which introduce $N\left( {N - 1} \right)/2$ independent thread bundles (or component flows) ${\vec v_{ij}}$. We only present one of these thread bundles in the figure, represented by a thick mauve line. Furthermore, the component flow fluxes should satisfy (\ref{fij}). Again, the green dashed lines represent RT surfaces.}
		\label{fig6}
	\end{center}	
\end{figure}

In this section, we will explicitly construct a model described by the multiflow language of bit threads, which will be called the ``holographic qubit threads model''. This model can be interpreted as an efficient ``decoding'' method of the entanglement structure of the boundary CFT state. More explicitly, it first ``distills'' the entanglement structure of the CFT state into the form of ``qubits'' (or bell pairs), and the information of these qubits is represented by the bit threads in a quantum superposition state. Then, from this set of bit threads, the SS states (which are the states in the Hilbert subspace of the holographic boundary CFT) of a specified set of extremal surfaces in bulk spacetime can be read and extracted, along with their entanglement relations with each other. We will specifically call these bit threads that can reflect the entanglement structure of the boundary quantum system to some extent and are endowed with the physical meaning of being in a quantum superposition state as qubit threads, to distinguish them from the general ``bit thread'' in the existing literature, which is merely used as a mathematical concept in convex programming theory on manifolds.

Specifically, we construct the qubit threads model as shown in figure~\ref{fig6}. Consider dividing the boundary quantum system into $N$ elementary regions denoted as ${A_1}$, ${A_2}$, ..., ${A_N}$. $N$ can be taken to be extremely large such that the size of each elementary region is very small, however, for physical reasons, we will make each small piece much larger than the Planck scale to ensure the validity of RT formula. The purpose of this division is to extract the information of the entanglement structure of the holographic boundary quantum system more sufficiently and precisely, so as to describe the mechanism of holographic principle more effectively. A natural way to do this is to take all the $N$ elementary regions to be of the same size (which is easy to do for the one-dimensional spatial space of a ${CFT_2}$). For the case involving $N$ elementary regions, $N\left( {N - 1} \right)/2$ independent thread bundles will be introduced. In other words, the multiflow characterizing the corresponding locking thread configuration will be composed of $N\left( {N - 1} \right)/2$ component flows ${\vec v_{ij}}$. Next, in order to for the constructed locking thread configuration to reflect the entanglement structure information of the boundary quantum system correctly and reasonably, we need to make use of the so-called ``CFF=PEE'' scheme proposed in~\cite{Lin:2021hqs}.  According to this scheme, a special type of locking bit thread configuration can be designed such that the component flow fluxes in the multiflow can exactly match the partial entanglement entropies in the boundary quantum system. In brief, this scheme requires that the fluxes $\left\{ {{F_{ij}}} \right\}$ of the locking thread configuration satisfy the following requirements:
\be\label{equ}{S_{a\left( {a + 1} \right) \ldots b}} = \sum\limits_{s,t} {{F_{st}}} , \,\,\,\,\, s \in \left\{ {a,a + 1, \cdots ,b} \right\}, t \notin \left\{ {a,a + 1, \cdots ,b} \right\},\ee
where ${S_{a\left( {a + 1} \right) \ldots b}}$ represents the entanglement entropy ${S_R}$ of a connected composite region $R = {A_{a\left( {a + 1} \right) \cdots b}} \equiv {A_a} \cup {A_{a + 1}} \cup  \cdots  \cup {A_b}$. This formula can be intuitively understood as that the entanglement entropy between $R$ and its completement $\bar R$ comes from the sum of fluxes of all the thread bundles connecting the elementary regions ${A_s}$ within $R$ and the elementary regions ${A_t}$ within $\bar R$ at both ends respectively. The point is that the system of equations~(\ref{equ}) contains $N\left( {N - 1} \right)/2$ constraints in total, so the exact value of each ${F_{ij}}$ can be precisely determined.

It is worth emphasizing again the necessity for the ``qubit threads'' to satisfy the constraints on fluxes ${F_{ij}}$ given by equation (\ref{equ}). This is because essentially ${F_{ij}}$ can be understood as a generalization of the multipart case of mutual information $I$. In particular, in the case involving only three elementary regions, we have (\ref{inf}). Furthermore, according to this scheme, one can obtain the expression of PEE in terms of flow fluxes~\cite{Lin:2021hqs}:
\be{s_R}\left( {{A_s}} \right) = \sum\limits_t {{F_{st}}} ,\,\,\,\, s \in \left\{ {a,a + 1, \cdots ,b} \right\}, t \notin \left\{ {a,a + 1, \cdots ,b} \right\}.\ee
Denoting the region sandwiched between ${A_i}$ and ${A_j}$ as $\tilde A = {A_{\left( {i + 1} \right) \cdots \left( {j - 1} \right)}}$  (which is a composite region composed of a set of elementary regions), then one can solve (\ref{equ}) and obtain~\cite{Lin:2021hqs}, 
\be\label{fij}{F_{ij}} = \frac{1}{2}\left( {S\left( {\tilde A \cup {A_i}} \right) + S\left( {\tilde A \cup {A_j}} \right) - S\left( {\tilde A} \right) - S\left( {{A_i} \cup \tilde A \cup {A_j}} \right)} \right).\ee
It is not difficult to find that this expression is equivalent to the so-called PEE proposal proposed in~\cite{Kudler-Flam:2019oru,Wen:2018whg}. Moreover, in fact, PEE is referred to as conditional mutual information (CMI) in quantum information theory (up to a conventional 1/2 factor)~\cite{Czech:2015kbp,Czech:2015qta}, and (\ref{fij}) is just the definition of CMI.

Given the constraints on the thread fluxes that our model needs to satisfy, in the next subsection we will show that for a set of specified non-crossing regions as shown in figure~\ref{fig6}, a locking bit thread configuration satisfying these constraints can always be explicitly constructed. The proof of the existence of this kind of locking bit thread configuration is an improvement of bulk cell decomposition method proposed in~\cite{Headrick:2020gyq}, and it adopts the recursive method. One interesting characteristic of the constructed thread configurations is that the threads in each thread bundle is uniformly distributed at the elementary regions that the threads connect and at the RT surfaces that the threads pass through. 

Finally, we obtain our locking qubit thread configuration model, which can lock a set of non-crossing boundary subregions simultaneously:~\footnote{Of course, this choice is just a convenience. We could have chosen another symmetrical set of non-crossing boundary subregions to be locked simultaneously. For example:
	$ I = \left\{ {{A_1}} \right\} \cup \left\{ {{A_2},{A_{12}}} \right\} \cup \left\{ {{A_3},{A_{123}}} \right\} \cup \left\{ {{A_4},{A_{1234}}} \right\} \cup  \cdots  \cup \left\{ {{A_N},{A_{12 \cdots N}}} \right\}$.}

\begin{scriptsize}\be\label{set}I = \left\{ {{A_1},{A_2}, \cdots ,{A_N}} \right\} \cup \left\{ {{A_{12}},{A_{34}}, \cdots ,{A_{\left( {N - 1} \right)N}}} \right\} \cup \left\{ {{A_{1234}},{A_{5678}}, \cdots ,{A_{\left( {N - 3} \right)\left( {N - 2} \right)\left( {N - 1} \right)N}}} \right\} \cup  \cdots  \cup \left\{ {{A_{123 \cdots N/2}},{A_{\left( {N/2 + 1} \right) \cdots N}}} \right\},\ee\end{scriptsize}
where we have used shorthand, e.g., ${A_{12}} \equiv {A_1} \cup {A_2}$, ${A_{123}} \equiv {A_1} \cup {A_2} \cup {A_3}$, etc, and we have chosen $N = {2^n}$ ($n$ is a positive integer) implicitly.

According to the thread/state prescription, and following the previous notation, then one can write out the density matrix of each RT surface directly and concretely. For example, the density matrix of surface ${\gamma _1}$ can be written as follows:
\begin{tiny}\be{\rho _{{\gamma _1}}} = \sum\limits_{{q_{12}} = 0}^{{2^{{F_{12}}}} - 1} {\sum\limits_{{q_{13}} = 0}^{{2^{{F_{13}}}} - 1} { \cdots \sum\limits_{{q_{1N}} = 0}^{{2^{{F_{1N}}}} - 1} {\frac{1}{{{2^{{S_1}}}}}\left( {\left| {{\phi _{12}} = {q_{12}}} \right\rangle  \otimes \left| {{\phi _{13}} = {q_{13}}} \right\rangle  \otimes  \cdots  \otimes \left| {{\phi _{1N}} = {q_{1N}}} \right\rangle } \right)\left( {\left\langle {{\phi _{12}} = {q_{12}}} \right| \otimes \left\langle {{\phi _{13}} = {q_{13}}} \right| \otimes  \cdots  \otimes \left\langle {{\phi _{1N}} = {q_{1N}}} \right|} \right)} } }, \ee\end{tiny}
i.e.,
\be\label{firi}{\rho _{{\gamma _1}}} = \sum\limits_{t \ne 1} {\sum\limits_{{q_{1t}} = 0}^{{2^{{F_{1t}}}} - 1} {\frac{1}{{{2^{{S_1}}}}}\left( {\prod\limits_{t \ne 1} {\left| {{\phi _{1t}} = {q_{1t}}} \right\rangle } } \right)\left( {\prod\limits_{t \ne 1} {\left\langle {{\phi _{1t}} = {q_{1t}}} \right|} } \right)} }. \ee
Similarly, for a more general connected boundary composite region $R = {A_{a\left( {a + 1} \right) \cdots b}} \equiv {A_a} \cup {A_{a + 1}} \cup  \cdots  \cup {A_b}$, the SS state of its corresponding holographic RT surface ${\gamma _{a\left( {a + 1} \right) \cdots b}}$ can be expressed as follows:
\be{\rho _{{\gamma _{a\left( {a + 1} \right) \cdots b}}}} = \sum\limits_{s,t} {\sum\limits_{{q_{st}} = 0}^{{2^{{F_{st}}}} - 1} {\frac{1}{{{2^{{S_{a\left( {a + 1} \right) \cdots b}}}}}}\left( {\prod\limits_{s,t} {\left| {{\phi _{st}} = {q_{st}}} \right\rangle } } \right)\left( {\prod\limits_{s,t} {\left\langle {{\phi _{st}} = {q_{st}}} \right|} } \right)} }, \ee
where $s \in \left\{ {a,a + 1, \cdots ,b} \right\}$, $t \notin \left\{ {a,a + 1, \cdots ,b} \right\}$.
In particular, similar to (\ref{sym}), we can also write out the  pure state corresponding to the closed surface that consists of all the RT surfaces of the first layer:
\be\left| \Psi  \right\rangle  = \sum\limits_{i,j} {\sum\limits_{{q_{ij}} = 0}^{{2^{{F_{ij}}}} - 1} {\frac{1}{{\sqrt {{2^{\sum\limits_{i,j} {{F_{ij}}} }}} }}\prod\limits_{k = 1}^N {\left( {\prod\limits_{\mathop {l,}\limits_{l \ne k} } {\left| {{\phi _{kl}} = {q_{kl}}} \right\rangle } } \right)} } } .\ee

On the other hand, to express that the state of the boundary quantum system $\left| {{\Psi ^{CFT}}} \right\rangle$ is effectively mapped to (or approximated to, with high accuracy) the state $\left| \Psi  \right\rangle$ described by the qubit threads through a process of entanglement distillation, formally we can write
\be\left| {{\Psi ^{CFT}}} \right\rangle  = {V^{ - 1}}\left| \Psi  \right\rangle, \ee
therefore
\be\left| \Psi  \right\rangle  = V\left| {{\Psi ^{CFT}}} \right\rangle. \ee
Then the isometry tensor $V$ represents the entanglement distillation process.
We can also define an isomorphic mapping between a single elementary region and its corresponding RT surface:
\be{V_i}:\;{\rho _{{A_i}}} \mapsto {\rho _{{\gamma _i}}},\ee
where ${V_i}$ is understood as OSED (one-shot entanglement distillation) in literature~\cite{Bao:2018pvs,Bao:2019fpq,Lin:2020ufd}.

The reader may wonder how, in this model, for subregions like ${A_{23}}$, ${A_{123}}$, etc., the information of the corresponding holographic RT surfaces can be ``read" out? Indeed, due to the nontrivial density bound that the bit threads must satisfy in the bulk, one cannot always simultaneously lock an arbitrary set of boundary subregions. However, we can always design a new non-crossing set without changing the requirements (\ref{equ}) imposed on thread fluxes. For example, we can do a ``rotation" on the set (\ref{set}), which leads to

\begin{scriptsize}\be I = \left\{ {{A_1},{A_2}, \cdots ,{A_N}} \right\} \cup \left\{ {{A_{23}},{A_{45}}, \cdots ,{A_{N1}}} \right\} \cup \left\{ {{A_{2345}},{A_{6789}}, \cdots ,{A_{\left( {N - 2} \right)\left( {N - 1} \right)N1}}} \right\} \cup  \cdots  \cup \left\{ {{A_{23 \cdots \left( {N/2 + 1} \right)}},{A_{\left( {N/2 + 2} \right) \cdots N1}}} \right\}.\ee\end{scriptsize}
In this way, we can get another locking qubit thread configuration, which can be used as a decoding mechanism to read out a new set of SS states of a series of bulk extremal surfaces and their entanglement relationships between each other from the entanglement structure information of the boundary CFT state.

%%%%%%%%%%%%%%%%%%%%%%%%%%%%%%%%%%%%%%%%%%%%%%%%%%%%%%%%%%%%%%%%%%%%%%
\subsection{Constructing holographic qubit threads model from locking bit thread configurations}\label{subsec4.2}
%%%%%%%%%%%%%%%%%%%%%%%%%%%%%%%%%%%%%%%%%%%%%%%%%%%%%%%%%%%%%%%%%%%

In this section, we will adopt an iterative approach, which can be called bulk cell gluing method~\cite{Headrick:2020gyq,Lin:2020yzf}, combined with the convex duality technique from the theory of convex optimization to prove that there is always a bit thread configuration that can lock any set of specified non-crossing regions and satisfy the ``CFF=PEE'' constraint at the same time, such that it can be used as a holographic qubit threads model. 
\begin{figure}[htbp]     \begin{center}
		\includegraphics[height=5cm,clip]{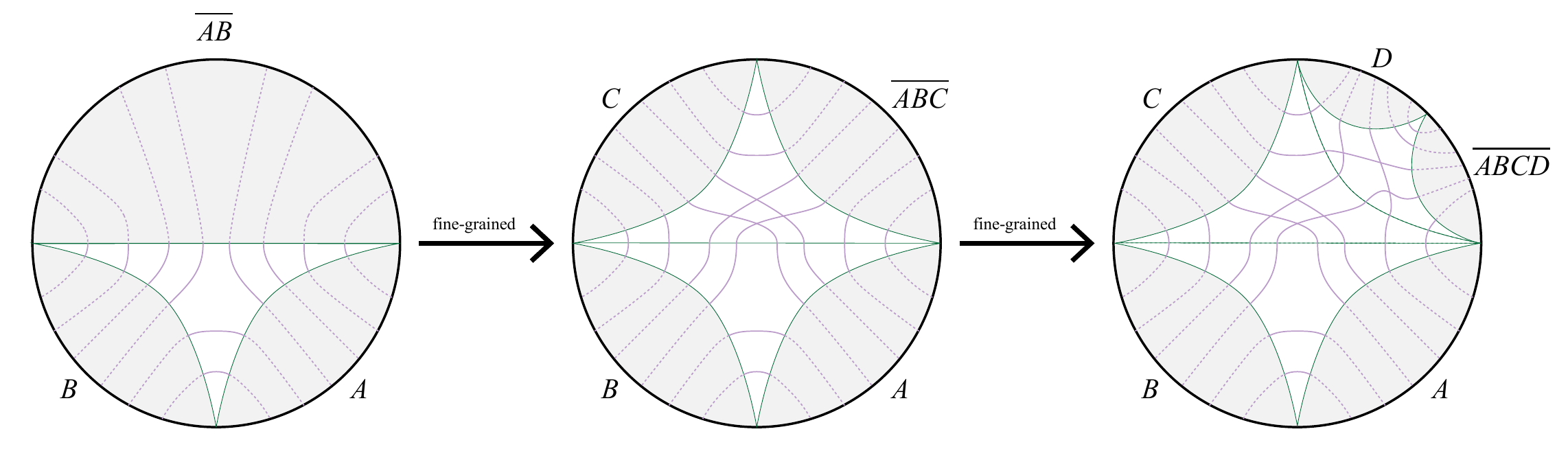}
		\caption{Given an initial bit thread configuration $V$ locking three regions $\{ A,B,AB\}$ simultaneously, by further decomposing a two-sided cell (grey region) into a three-sided cell (white region) and two smaller two-sided cells, one can use the replacing method to construct a new fine-grained thread configuration. This refinement process can be repeated iteratively. This ``replacement" description is equivalent to the bulk cell gluing method in Section~\ref{subsec2.1}.}
		\label{fig421}
	\end{center}	
\end{figure}

First of all, as shown in figure~\ref{fig421}, note that a thread configuration that can lock a specified set of boundary subregions $\{ A,B,AB, \cdots \}$ must equivalently lock its corresponding set of RT surfaces $\{ {\gamma _A},{\gamma _B},{\gamma _{AB}}, \cdots \} $ (due to the divergenceless property of bit threads), therefore, we can focus on the latter. A set of non-crossing boundary subregions corresponds to a set of non-intersecting RT surfaces. These RT surfaces divide the bulk spacetime into small cells, whose boundaries are RT surfaces or boundary subregions. As shown in the figure, for the sake of discussion, let us focus on the case where the bulk spacetime is decomposed into only two kinds of basic cells. One is the three-sided cell whose boundary is composed of three adjacent RT surfaces, and the other is the two-sided cell whose boundary is composed of an RT surface and a corresponding boundary subregion. Note that other types of cells containing more sides can always return to the case we considered by adding more auxiliary RT surfaces~\cite{Lin:2020yzf}.

Now we use the idea of recursion method. As shown in figure~\ref{fig421}, suppose at the beginning we have found a bit thread configuration $V$ that can lock three regions $\{ A,B,AB\}$ simultaneously, or equivalently, lock $\{ {\gamma _A},{\gamma _B},{\gamma _{AB}}\} $. In addition, we label the three-sided cell enclosed by ${\gamma _A}$, ${\gamma _B}$, ${\gamma _{AB}}$ as $\mathcal{C}\left( {AB} \right)$, and the three two-sided cells as $\mathcal{W}\left( A \right)$, $\mathcal{W}\left( B \right)$, and $\mathcal{W}\left( {\overline {AB} } \right)$ (because they are entanglement wedges of $A$, $B$, and $\overline {AB}$, respectively). It can be observed that such a locking bit thread configuration has three types of thread bundles traveling in different directions in each three-sided cell (painted in white in the figure), and each type connects a pair of RT surfaces, while in each two-sided cell (painted in light gray), the threads simply travel in one direction, connecting the boundary subregion to the corresponding RT surface. Next, we can further decompose the two-sided cell surrounded by ${\gamma _{AB}}$ and $\overline {AB} $ into a three-sided cell and two two-sided cells, and we are considering a corresponding problem: can we find a (new) bit thread configuration that can further lock the set of boundary subregions $\{ A,B,AB,C,ABC\} $ simultaneously (or equivalently, lock $\{ {\gamma _A},{\gamma _B},{\gamma _{AB}},{\gamma _C},{\gamma _{ABC}}\} $). The idea is that we can replace only the part of the bit threads inside $\mathcal{W}\left( {\overline {AB} } \right)$ based on the original locking bit thread configuration $V$~\cite{Headrick:2020gyq,Lin:2020yzf}. This process can be referred to as the fine-graining of the bit thread configuration, since it can be accordingly understood as a further fine-graining of the entanglement structure of the boundary regions. One can see that by such a refinement process, the bit threads in $\mathcal{W}\left( {\overline {AB} } \right)$ that originally only extended in one direction were replaced by a thread configuration that have different types of thread bundles along different directions and might lead to intersecting threads (due to the requirement of ``CFF=PEE''). In other words, more entanglement information was presented. Conversely, the original partial thread configurations in two-sided cells present merely coarse-graining information. Therefore, one can ask, can we always find such a more refined partial thread configuration (to replace the original one) satisfying the constraint of bit threads and ``PEE=CFF'' condition? Next we will use an explicit construction to show that the answer is positive. Once the proof is complete, as shown in figure~\ref{fig421}, we can continue to iterate this refinement process, repeatedly subdividing the two-sided cells (gray areas in the figure) into three-sided cells (white areas) and smaller two-sided cells. Finally, we can obtain an exquisite locking bit thread configuration that can lock a large $N$ number of elementary regions and a set of non-crossing composite regions simultaneously while satisfy the ``CFF=PEE'' constraint at the same time.

The description of the ``replacement" of partial thread configuration here can also be formulated as the bulk cell gluing method proposed in~\cite{Headrick:2020gyq}(as we reviewed in Section~\ref{subsec2.1}, see also~\cite{Lin:2020yzf}). Assume that we have found a thread configuration in $\mathcal{W}\left( T \right)$ that can lock $n$ adjacent elementary regions $t$ (denoted as $t$=1, 2, ... , $n$) making up $T$, as well as a specified set of non-crossing composite regions in $T$, and, in particular, the region $T$ itself. Note that locking $T$ also equivalently means locking its corresponding RT surface $\tilde T$ or locking its complement $\bar T$. Then we divide $\bar T$ into two new elementary regions, $R$ and $S$. What we need to do is to find a thread configuration for the cell spacetime $\mathcal{W}\left( {RS} \right)$ that can further lock $\left\{ {R,S,\bar T = RS} \right\}$, or its corresponding set of RT surfaces $\left\{ {\tilde R,\tilde S,\tilde T} \right\}$ simultaneously. The bulk cell gluing method points out that, due to the max flow-min cut theorem, on the minimal surface $\tilde T$, the bit threads must be orthogonal to $\tilde T$ and the thread density exactly reaches the maximum value $\frac{1}{{4{G_N}}}$, therefore, the locking bit thread configurations in two submanifolds $\mathcal{W}\left( T \right)$ and $\mathcal{W}\left( {RS} \right)$ can just be glued together at the minimal surface $\tilde T$ due to continuity, and thus providing an overall locking bit thread configuration of the whole spacetime. In~\cite{Headrick:2020gyq}, this gluing process can be achieved unambiguously simply by constructing a special thread configuration described by a $flow$ (which is a special case of the $multiflow$) wherein the bit threads are locally parallel everywhere. However, due to the new non-trivial ``CFF=PEE'' constraint introduced by the holographic qubit threads model, simply using flow to implement gluing will not work in general. The reason is straightforward. In a flow configuration, by definition, the bit threads cannot cross each other. As a result, in the case involving the gluing of a sufficient number of cells, the threads starting from some boundary elementary region may not be able to reach some other distant elementary regions. In other words, this will inevitably make some of the CFFs (component flow fluxes) vanish and possibly violate the ``CFF=PEE'' condition, which is designed to ensure the bit threads to faithfully represent the entanglement structure of the boundary quantum system. Therefore, we need to propose a more detailed gluing scheme. The key point is to make sure that the bit threads starting from one elementary region can reach any other elementary region at the end. 

In the following, we will show that the desired gluing can be achieved by requiring that each thread bundle (or component flow) be uniformly distributed everywhere on all the RT surfaces through which it passes. Furthermore, fortunately, using the strong duality technique of convex programs, we can prove that there always exists a locking bit thread configuration meeting this requirement of uniform distribution. Given our uniform scheme, we get the following story: there are $n$ component flows ${v_{t\tilde T}}$ flowing into the $\tilde T$ surface, and each component flow is uniformly distributed on it. Due to the locking requirement, the number of perpendicularly intersecting threads on each unit area of $\tilde T$ surface is $\frac{1}{{4{G_N}}}$, among which there are $n$ kinds of threads (belonging to different independent component flows ${v_{t\tilde T}}$), whose number is denoted as a constant ${c_t}$ respectively, i.e.,
\be{\left. {\sum\limits_t {\rho \left( {{{\vec v}_{t\tilde T}}} \right)} } \right|_{\tilde T}} = \frac{1}{{4{G_N}}},\ee
where $\rho $ represents thread density, which is defined as the sum of the lengths of threads per unit volume~\footnote{or the maximal number of threads across a unit disk~\cite{Headrick:2020gyq}. Actually the two definitions are technically subtly different, see the discussion in~\cite{Headrick:2020gyq}.}, i.e.,
\be\rho \left( {{{\vec v}_{t\tilde T}}} \right) = \left| {{{\vec v}_{t\tilde T}}} \right|.\ee
The numbers of threads in the thread bundle are described by the fluxes of the component flows ${v_{t\tilde T}}$:
\be{F_{t\tilde T}} = \int_{\tilde T} {{{\vec v}_{t\tilde T}}}  = \int_{\tilde S + \tilde R} {{{\vec v}_{t\tilde T}}} ,\ee
thus
\be{c_t} = {\left. {\rho \left( {{{\vec v}_{t\tilde T}}} \right)} \right|_{\tilde T}} = \frac{{{F_{t\tilde T}}}}{{Area\left( {\tilde T} \right)}}.\ee
Obviously, this satisfies
\be\sum\limits_t {{c_t}}  = \frac{1}{{4{G_N}}}.\ee
Next, to ensure the continuity of gluing under the CFF=PEE constraint, we split each ${\vec v_{t\tilde T}}$ thread bundle reaching the $\tilde T$ surface into two independent thread bundles:
\be{\vec v_{t\tilde T}} = {\vec v_{tS}} + {\vec v_{tR}}.\ee
Specifically, we decompose the constant ${c_t}$ into two new constants,
\be{c_t} = {c_{tS}} + {c_{tR}}.\ee
Since the ${\vec v_{t\tilde T}}$ thread bundle is uniformly distributed everywhere on the $\tilde T$ surface, we can always unambiguously assign ${c_{tS}}$ threads in each unit area such that there are ${c_{tS}}$ threads entering the $S$ surface (and renamed as ${\vec v_{tS}}$ flow), while ${c_{tR}}$ threads entering the $R$ surface (and renamed as ${\vec v_{tR}}$ flow). The CFF=PEE conditions require:
\be{F_{t\tilde T}} = {F_{tS}} + {F_{tR}},\ee
which determine
\be\label{cts}{c_{tS}} = \frac{{{F_{tS}}}}{{Area\left( {\tilde T} \right)}},\quad {c_{tR}} = \frac{{{F_{tR}}}}{{Area\left( {\tilde T} \right)}}.\ee
We do the assignment for every thread bundle ${\vec v_{t\tilde T}}$ on $\tilde T$, furthermore, in order to apply the recursion method, the shunted thread bundles are also required to be uniformly distributed everywhere when it reaches $\tilde S$ surface and $\tilde R$ surface, and the densities are denoted as constant ${c'_{tS}}$, ${c'_{tR}} $ respectively:
\be\label{cts2}{c'_{tS}} = \frac{{{F_{tS}}}}{{Area\left( {\tilde S} \right)}},\quad {c'_{tR}} = \frac{{{F_{tR}}}}{{Area\left( {\tilde R} \right)}}\ee
Of course, as analyzed above, the refinement process of bit thread configuration will also introduce a new thread bundle ${\vec v_{RS}}$ connecting $R$ and $S$. We also require this thread bundle to be uniformly distributed everywhere on the $\tilde S$ surface and $\tilde R$ surface it passes through, wherein the density is denoted as constant $\alpha $ and $\beta $, respectively, i.e.,
\be\label{alp}\alpha  = \frac{{{F_{RS}}}}{{Area\left( {\tilde S} \right)}},\quad \beta  = \frac{{{F_{RS}}}}{{Area\left( {\tilde R} \right)}}\ee
Then according to the CFF=PEE constraint, on the three RT surfaces $\tilde T$, $\tilde S$ and $\tilde R$ we have respectively 
\be\label{sch}\begin{array}{l}
	\frac{1}{{4{G_N}}} = \sum\limits_t {{c_{tS}}}  + \sum\limits_t {{c_{tR}}} \\
	\frac{1}{{4{G_N}}} = \sum\limits_t {{{c'}_{tS}}}  + \alpha \\
	\frac{1}{{4{G_N}}} = \sum\limits_t {{{c'}_{tR}}}  + \beta 
\end{array}.\ee
In a word, given the desired correct CFF=PEE condition, we can always design a feasible gluing scheme according to~(\ref{sch}).

Next we turn our attention to a single key step of the recursion method. We need to prove that, considering $\mathcal{W}\left( {RS} \right) =\mathcal{ C}\left( {RS} \right) \cup \mathcal{W}\left( R \right) \cup \mathcal{W}\left( S \right)$ as an individual spacetime, there always exists a locking bit thread configuration that can simultaneously lock the three ``elementary regions" $\tilde T$, $S$ and $R$ (which compose its boundary), and meet the following conditions: It is composed of $2n+1$ component flows $\left\{ {{{\vec v}_{tS}}} \right\}$, $\left\{ {{{\vec v}_{tR}}} \right\}$, ${\vec v_{RS}}$, and these component flows meet the requirements (\ref{cts}), (\ref{cts2}), (\ref{alp}), i.e., their densities are constant at the three RT surfaces $\tilde T$, $\tilde S$ and $\tilde R$. In fact, this problem can be further simplified. We just need to show that, considering the three-sided cell $\mathcal{C}\left( {RS} \right)$ as an individual spacetime, there always exists a thread configuration described by a multiflow consisting of three component flows ${\vec v_{\tilde T\tilde S}}$, ${\vec v_{\tilde T\tilde R}}$, ${\vec v_{\tilde R\tilde S}}$ that can simultaneously lock the three elementary regions $\tilde T$, $\tilde S$, $\tilde R$ at the boundary of $\mathcal{C}\left( {RS} \right)$, while satisfy the additional boundary conditions that the component flow densities at the boundary must be constant, i.e.,
\be{\rm{on}}\;\tilde T:\quad \left| {{{\vec v}_{\tilde T\tilde S}}} \right| = {c_{TS}},\;\;\left| {{{\vec v}_{\tilde T\tilde R}}} \right| = {c_{TR}},\ee
\be{\rm{on}}\;\tilde S:\quad \left| {{{\vec v}_{\tilde T\tilde S}}} \right| = {{c'}_{TS}},\;\;\left| {{{\vec v}_{\tilde R\tilde S}}} \right| = \alpha, \ee
\be{\rm{on}}\;\tilde R:\quad \left| {{{\vec v}_{\tilde T\tilde R}}} \right| = {{c'}_{TR}},\;\;\left| {{{\vec v}_{\tilde R\tilde S}}} \right| = \beta. \ee
Then by defining
\be{\vec v_{tS}} = \frac{{{F_{tS}}}}{{\sum\limits_t {{F_{tS}}} }}{\vec v_{\tilde T\tilde S}} = \frac{{{F_{tS}}}}{{{F_{\tilde T\tilde S}}}}{\vec v_{\tilde T\tilde S}},\ee
\be{\vec v_{tR}} = \frac{{{F_{tR}}}}{{\sum\limits_t {{F_{tR}}} }}{\vec v_{\tilde T\tilde R}} = \frac{{{F_{tR}}}}{{{F_{\tilde T\tilde R}}}}{\vec v_{\tilde T\tilde R}},\ee
\be{\vec v_{RS}} = {\vec v_{\tilde R\tilde S}},\ee
we complete the construction, since the construction of the bit thread configurations in $\mathcal{W}\left( R \right)$ and $\mathcal{W}\left( S \right)$ are trivial (``coarse-grained"). In this way, we have reduced the original problem to a fairly simple subproblem involving only three elementary regions, with simple extra boundary conditions. By using the systematic method of convex dualization developed in~\cite{Headrick:2020gyq}, the existence of the solution to this convex optimization problem can be proved immediately. To be self-contained, we put the explicit proof in the appendix.

Once we complete this key step, the proof of the existence of the whole overall locking bit thread configuration is complete. By using the recursion method, we consider that initially $T$ contains only one elementary region. In this case, obviously there exists a trivial locking bit thread configuration that can lock the two boundaries $T$ and $\tilde T$ of the submanifold $\mathcal{W}\left( T \right)$ simultaneously, and satisfy the uniformly distributed requirement on the minimal surface $\tilde T$. Therefore, the proof works for the case $n=1$. Since according to our scheme, after each step of construction, the independent thread bundles are always uniformly distributed on the RT surfaces that ``segment'' the bulk, we can iteratively apply the above argument, and finally construct a desired locking bit thread configuration for a non-crossing set which involves an arbitrary number of elementary regions.

%%%%%%%%%%%%%%%%%%%%%%%%%%%%%%%%%%%%%%%%%%%%%%%%%%%%%%%%%%%%%%%%%%%%%%
\section{Relations with other holographic models}\label{sec5}
%%%%%%%%%%%%%%%%%%%%%%%%%%%%%%%%%%%%%%%%%%%%%%%%%%%%%%%%%%%%%%%%%%%%%%

%%%%%%%%%%%%%%%%%%%%%%%%%%%%%%%%%%%%%%%%%%%%%%%%%%%%%%%%%%%%%%%%%%%%%%
\subsection{Relation with holographic tensor network models: qubit threads make disentangling easy}\label{subsec5.1}
%%%%%%%%%%%%%%%%%%%%%%%%%%%%%%%%%%%%%%%%%%%%%%%%%%%%%%%%%%%%%%%%%%%%%%
\begin{figure}[htbp]     \begin{center}
		\includegraphics[height=8cm,clip]{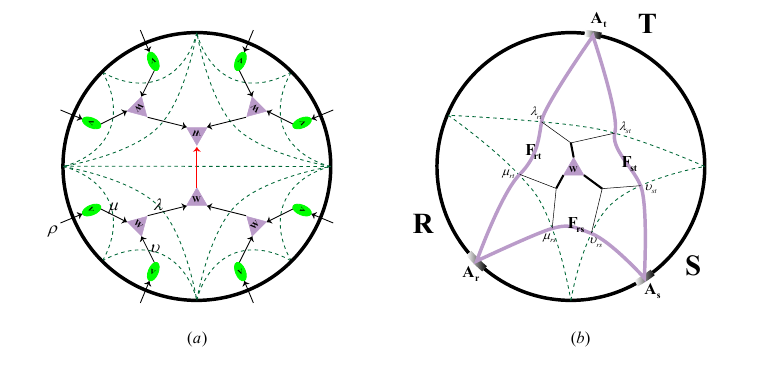}
		\caption{(a) To match our locking qubit thread model with a holographic tensor network model, using thread /state correspondence, two types of tensor $V$ and $W$ are constructed. $V$ represents the distillation tensor, while $W$ plays the role of a $disentangler$. An inward-pointing arrow corresponds to a lower index, while an outward-pointing arrow corresponds to an upper index. (b) The thread/state prescription of the explicit expression for a general $W$ tensor.}
		\label{fig7}
	\end{center}	
\end{figure}

Our locking qubit thread configuration model is actually very similar to the holographic tensor network models.~\footnote{For a preliminary discussion on the connection between bit threads and holographic tensor networks, see~\cite{Lin:2020yzf}.} We first note that the two kinds of models describe the state of the holographic RT surface corresponding to a boundary subregion in a consistent manner. In the holographic tensor network models, the holographic RT surface corresponding to a (connected) boundary subsystem $R$ is defined as a minimal cut to the network such that the two halves of the network are totally adjacent to $R$ and $\bar R$, respectively~\cite{Swingle:2009bg,Swingle:2012wq}. Thus, the direct product of the states represented by the cut-off legs defines the state describing this RT surface. While in our model, we have transformed the language of geometric surfaces into the language of threads. Therefore, we can also describe the state of the holographic RT surface corresponding to a connected boundary subsystem $R$ (which is generally a composite region composed of a set of adjacent elementary regions) in a similar way, i.e., as a cut to the thread configuration such that the thread configuration is divided into two halves, each of which is entirely adjacent to $R$ and $\bar R$ respectively. Therefore, the direct product of the reduced states represented by the cut-off threads defines the state describing the holographic RT surface ${\gamma _R}$.

In fact, as hinted earlier, the threads in our model exactly play the roles of legs in the tensor network models, and both are endowed with a Hilbert space. But the difference is that a leg in the latter refers to the quantum correlation on a local point, while our thread refers to the quantum correlation between a pair of points separated by some distance. It turns out that, this characteristic of qubit threads can intuitively describe an important concept in holographic tensor network models (especially MERA-like tensor networks), namely, $disentangler$~\cite{Vidal:2007hda,Vidal:2008zz,Swingle:2009bg,Swingle:2012wq}. The point is that the threads that define ${\gamma _R}$ are merely those connecting $R$ to $\bar R$, therefore, the inner threads that represent the internal entanglement within $R$ do not contribute to the entanglement entropy of $R$. The former represents long range entanglement, and the latter represents short range entanglement.

Next we explicitly present an explicit matching between our locking qubit thread configuration model and the holographic tensor network models. More precisely, we will show that the thread configuration in each of these small bulk cells formed by the segmentation of the RT surfaces can be characterized in the form of a tensor. A preliminary discussion on this issue has been present in~\cite{Lin:2020yzf}, where the bit-thread representation of a kind of holographic tensor network has been preliminarily discussed (see also~\cite{Lin:2020ufd}).

For simplicity, we draw a schematic by choosing $N$ equal to 8, see figure~\ref{fig7}(a). However, remember that, actually $N$ should be taken as a very large number. Anyway, this simple schematic is enough to help us find the universal expression directly. In order to directly match our locking qubit thread model with a holographic tensor network model, we can construct two kinds of tensors using thread /state correspondence, denoted as $V$ and $W$ in the figure respectively. Wherein $V$ represents the distillation tensor, which is defined as mapping the reduced state of each boundary elementary region to the SS state associated with its RT surface. In this paper, we do not focus on the study of $V$ tensor, which is merely a pro forma definition at the moment. The $W$ tensor plays the role of a $disentangler$, which maps the SS states of the two RT surfaces in the previous layer to the SS state of the RT surface in the next layer. In addition to representing an index of each tensor in the tensor network by a leg extending from a vertex, we also assign an arrow direction to each leg, such that the inward-pointing arrow corresponds to the lower index of the tensor, while the outward-pointing arrow corresponds to the upper index. We also adopt the convention that the arrows of the entire network point from the outermost layer to the innermost layer. Due to the symmetry of our configuration, the arrow direction of each leg in the whole network can be uniquely determined. According to this convention, each $V$ tensor has an arrow pointing inward and an arrow pointing outward, while each $W$ tensor has two pointing-inward arrows and an pointing-outward arrow, except at the last step, there's a $W$ tensor that all the arrows are pointing inward.

Anyway, we can first write the $W$ tensor as $W_{\mu \nu }^\lambda $ pro forma, where $\mu $ and $\nu $ represent the SS states of the two RT surfaces of the previous layer, respectively, while $\lambda $ represents the SS state of the RT surface of the inner layer. Then the thread/state correspondence indicates that $\mu $, $\nu $ and $\lambda $ could be further decomposed into smaller indices. To see this, taking $W_{{\mu _1}{\nu _2}}^{{\lambda _{\left\{ {12} \right\}}}}$, which denotes the $W$ tensor connecting ${\gamma _1}$, ${\gamma _2}$ with ${\gamma _{\left\{ {12} \right\}}}$ in the schematic diagram, as an example, we first write out the pure state corresponds to the closed surface ${\gamma _1} \cup {\gamma _2} \cup {\gamma _{\left\{ {12} \right\}}}$, which is the generalized expression of (\ref{sym}) :
\begin{footnotesize}\be\label{wstate} \begin{array}{l}
	\left| {{\Psi _{{\gamma _1} \cup {\gamma _2} \cup {\gamma _{\left\{ {12} \right\}}}}}} \right\rangle  = \left( {\sum\limits_{{q_{13}} = 0}^{{2^{{F_{13}} - 1}}} {\sum\limits_{{q_{14}} = 0}^{{2^{{F_{14}} - 1}}} { \cdots \sum\limits_{{q_{1N}} = 0}^{{2^{{F_{1N}} - 1}}} {} } } } \right)\left( {\sum\limits_{{q_{23}} = 0}^{{2^{{F_{23}} - 1}}} {\sum\limits_{{q_{24}} = 0}^{{2^{{F_{24}} - 1}}} { \cdots \sum\limits_{{q_{2N}} = 0}^{{2^{{F_{2N}} - 1}}} {} } } } \right)\sum\limits_{{q_{12}} = 0}^{{2^{{F_{12}} - 1}}} {\frac{1}{{\sqrt {{2^{\left( {{F_{13}} + {F_{14}} +  \cdots {F_{1N}}} \right) + \left( {{F_{23}} + {F_{24}} +  \cdots  + {F_{2N}}} \right) + {F_{12}}}}} }}}  \cdot \\
	\left( {\left| {{\phi _{12}} = {q_{12}}} \right\rangle  \otimes \left| {{\phi _{13}} = {q_{13}}} \right\rangle  \otimes  \cdots  \otimes \left| {{\phi _{1N}} = {q_{1N}}} \right\rangle } \right)
	\otimes \left( {\left| {{\phi _{12}} = {q_{12}}} \right\rangle  \otimes \left| {{\phi _{23}} = {q_{23}}} \right\rangle  \otimes  \cdots  \otimes \left| {{\phi _{2N}} = {q_{2N}}} \right\rangle } \right)\\
	\otimes \left( {\left| {{\phi _{13}} = {q_{13}}} \right\rangle  \otimes \left| {{\phi _{14}} = {q_{14}}} \right\rangle  \otimes  \cdots  \otimes \left| {{\phi _{1N}} = {q_{1N}}} \right\rangle  \otimes \left| {{\phi _{23}} = {q_{23}}} \right\rangle  \otimes \left| {{\phi _{24}} = {q_{24}}} \right\rangle  \otimes  \cdots  \otimes \left| {{\phi _{2N}} = {q_{2N}}} \right\rangle } \right)
\end{array}.\ee\end{footnotesize}
According to (\ref{wstate}), we can write
\be\label{tensor}W_{{\mu _1}{\nu _2}}^{{\lambda _{\left\{ {12} \right\}}}} = W_{{\mu _{12}}{\mu _{13}} \cdots {\mu _{1N}}{\nu _{21}}{\nu _{23}} \cdots {\nu _{2N}}}^{{\lambda _{13}}{\lambda _{14}} \cdots {\lambda _{1N}}{\lambda _{23}}{\lambda _{24}} \cdots {\lambda _{2N}}},\ee
where the indices ${\lambda _{13}}$, ${\lambda _{14}}$, . ,${\lambda _{1N}}$, ${\lambda _{23}}$, ${\lambda _{24}}$, ..., ${\lambda _{2N}}$ represent the state spaces of the SS bits, which are respectively determined by the states of  ${\vec v_{13}}$, ${\vec v_{14}}$, ..., ${\vec v_{1N}}$, ${\vec v_{23}}$, ${\vec v_{24}}$, ..., ${\vec v_{2N}}$, i.e., the thread bundles intersecting with surface ${\gamma _{\left\{ {12} \right\}}}$, and similarly the indices $ {\mu _{12}}$, ${\mu _{13}}$, ..., ${\mu _{1N}}$ represent the state spaces of the SS bits, which are respectively determined by the states of  ${\vec v_{12}}$, ${\vec v_{13}}$, ..., ${\vec v_{1N}}$, i.e., the thread bundles intersecting with surface ${\gamma _1}$, while the indices ${\nu _{21}}$, ${\nu _{23}}$, ..., ${\nu _{2N}}$ represent the state spaces of the SS bits, which are respectively determined by the states of  ${\vec v_{21}}$, ${\vec v_{23}}$, ..., ${\vec v_{2N}}$, i.e., the thread bundles intersecting with surface $ {\gamma _2}$. Our correspondence rules state that the value ranges of these indices ${\lambda _{ij}}$, ${\mu _{ij}}$, ${\nu _{ij}}$ are all possible states of the thread bundles ${\vec v _{ij}}$, i.e.,
\be\left\{ {{q_{ij}}\left| {{q_{ij}} \in N,0 \le {q_{ij}} \le {2^{{F_{ij}}}} - 1} \right.} \right\}.\ee
It can be read from (\ref{wstate}) that each component of the $W$ tensor follows such a rule: the value of those components conforming to ${\lambda _{ij}} = {\mu _{ij}} = {\nu _{ij}}$ is equal to $\frac{1}{{\sqrt {{2^{\left( {{F_{13}} + {F_{14}} +  \cdots {F_{1N}}} \right) + \left( {{F_{23}} + {F_{24}} +  \cdots  + {F_{2N}}} \right) + {F_{12}}}}} }}$, while all the other components are equal to 0, that is,
\be\label{wtensor} W_{{\mu _1}{\nu _2}}^{{\lambda _{\left\{ {12} \right\}}}} = W_{{\mu _{12}}{\mu _{13}} \cdots {\mu _{1N}}{\nu _{21}}{\nu _{23}} \cdots {\nu _{2N}}}^{{\lambda _{13}}{\lambda _{14}} \cdots {\lambda _{1N}}{\lambda _{23}}{\lambda _{24}} \cdots {\lambda _{2N}}} = \left\{ {\begin{array}{*{20}{c}}
		{\frac{1}{{\sqrt {{2^{\left( {{F_{13}} + {F_{14}} +  \cdots {F_{1N}}} \right) + \left( {{F_{23}} + {F_{24}} +  \cdots  + {F_{2N}}} \right) + {F_{12}}}}} }},\quad {\lambda _{ij}} = {\mu _{ij}} = {\nu _{ij}}}\\
		{0.\quad {\rm{otherwise}}}
\end{array}} \right.\ee
To say that (\ref{wstate}) is a normalized pure state is equivalent to saying that the $W$ tensor (\ref{wtensor}) is a unitary tensor.

Seeing figure~\ref{fig7}(b), it is not difficult to write a universal expression for the $W$ tensor of any layer immediately. Again, we first write down the corresponding pure state of the closed surface that is the boundary of the cell of each layer. Consider a cell consisting of three RT surfaces ${\gamma _R}$, ${\gamma _S}$ and ${\gamma _T}$ corresponding to three adjacent composite boundary regions $R$, $S$ and $T$, respectively. Let us denote 
\be\begin{array}{l}
	R =  \cup {A_r},\quad r \in {\mathcal{R}}\\
	S =  \cup {A_s},\quad s \in {\mathcal{S}}\\
	T =  \cup {A_t},\quad t \in {\mathcal{T}},
\end{array}\ee
which satisfy $R \cup S \cup T = M$, that is, $R$, $S$, $T$ exactly compose the whole boundary system $M$. And we specially denote the elementary regions within $R$, $S$ and $T$ as ${A_r}$, ${A_s}$ and ${A_t}$ respectively, or in other words, we successively (in counterclockwise direction) refer to the indices $i$ of the elementary regions ${A_i}$ in $R$, $S$ and $T$ as $r$, $s$, and $t$ respectively. For convenience, the sets of indices $r$, $s$, and $t$ are denoted as ${\mathcal{R}}$, ${\mathcal{S}}$, ${\mathcal{T}}$ respectively. Following these notations, we can write:
\begin{tiny}\be\left| {{\Psi _{{\gamma _R} \cup {\gamma _S} \cup {\gamma _T}}}} \right\rangle  = \left( {\sum\limits_{r,t} {\sum\limits_{{q_{rt}} = 0}^{{2^{{F_{rt}}}} - 1} {} } } \right)\left( {\sum\limits_{s,t} {\sum\limits_{{q_{st}} = 0}^{{2^{{F_{st}}}} - 1} {} } } \right)\left( {\sum\limits_{r,s} {\sum\limits_{{q_{rs}} = 0}^{{2^{{F_{rs}}}} - 1} {} } } \right)\frac{1}{{\sqrt {{2^{\sum\limits_{r,t} {{F_{rt}}}  + \sum\limits_{s,t} {{F_{st}}}  + \sum\limits_{r,s} {{F_{rs}}} }}} }}\left( {\prod\limits_{r,t} {\left| {{\phi _{rt}} = {q_{rt}}} \right\rangle } } \right) \otimes \left( {\prod\limits_{s,t} {\left| {{\phi _{st}} = {q_{st}}} \right\rangle } } \right) \otimes \left( {\prod\limits_{r,s} {\left| {{\phi _{rs}} = {q_{rs}}} \right\rangle } } \right),\ee\end{tiny}
then similarly we obtain
\be\label{w} W_{{\mu _R}{\nu _S}}^{{\lambda _T}} = W_{\left\{ {{\mu _{rt}}} \right\}\left\{ {{\mu _{rs}}} \right\}\left\{ {{\nu _{rs}}} \right\}\left\{ {{\nu _{st}}} \right\}}^{\left\{ {{\lambda _{rt}}} \right\}\left\{ {{\lambda _{st}}} \right\}} = \left\{ {\begin{array}{*{20}{c}}
		{\frac{1}{{\sqrt {{2^{\sum\limits_{r,t} {{F_{rt}}}  + \sum\limits_{s,t} {{F_{st}}}  + \sum\limits_{r,s} {{F_{rs}}} }}} }},\quad {\lambda _{ij}} = {\mu _{ij}} = {\nu _{ij}}}\\
		{0.\quad otherwise}
\end{array}} \right.\ee
And (\ref{wstate}) and (\ref{wtensor}) are the special results of the simplest case where we set $R = {A_1}$, $S = {A_2}$, $T = {A_3} \cup {A_4} \cup  \cdots  \cup {A_N}$ in this general case.

From the expression of (\ref{w}), it is clear that the lower index of the $W$ tensor has the $rs$ component representing the entanglement between a pair of elementary regions ${A_r}$ and ${A_s}$, while the upper index has no $rs$ component. For example, it can be seen from (\ref{wtensor}) that the upper index of the $W$ tensor does not contain $12$ components, which indicates that the $W$ tensor is mapping a state containing the entanglement between ${A_1}$ and ${A_2}$ to a state absence of this part of entanglement. Therefore, the $W$ tensor is exactly playing the role of a $disentangler$ (of course, it also plays the role of the $coarse$-$grainer$ at the same time), because it is such a unitary operation that just converts the state of two entangled blocks in the previous layer into a state of a block without internal entanglement. This process of disentanglement can be intuitively understood as removing the contribution of the inner threads that represent the short-range entanglement! This process is carried out successively, exactly in line with the idea of MERA-like tensor network, that is, the effect of short-range entanglement is successively removed and only the effect of long-range entanglement is finally focused on.

On the other hand, we can directly write the distillation tensor $V$ pro forma as
\be V_{{\rho _i}}^{{\mu _i}} = V_{{\rho _i}}^{\left\{ {{\mu _{ij}}} \right\}},\ee
where $j$ runs from 1 to $N$ (except for $j$ itself), and the lower index ${\rho _i}$ represents the reduced state of the elementary region ${A_i}$, while ${\mu _i}$ represents the SS state of the RT surface ${\gamma _i}$ corresponding to ${A_i}$. For example, in figure~\ref{fig7}(a), for region ${A_1}$, 
\be V_{{\rho _1}}^{{\mu _1}} = V_{{\rho _1}}^{{\mu _{12}}{\mu _{13}} \cdots {\mu _{1N}}}.\ee
The $V$ tensors are isometries~\cite{Bao:2018pvs,Bao:2019fpq,Lin:2020ufd}.

%%%%%%%%%%%%%%%%%%%%%%%%%%%%%%%%%%%%%%%%%%%%%%%%%%%%%%%%%%%%%%%%%%%%%%
\subsection{Relation with kinematic space: qubit threads as CMI threads}\label{subsec5.2}
%%%%%%%%%%%%%%%%%%%%%%%%%%%%%%%%%%%%%%%%%%%%%%%%%%%%%%%%%%%%%%%%%%%%%%
\begin{figure}[htbp]     \begin{center}
		\includegraphics[height=7cm,clip]{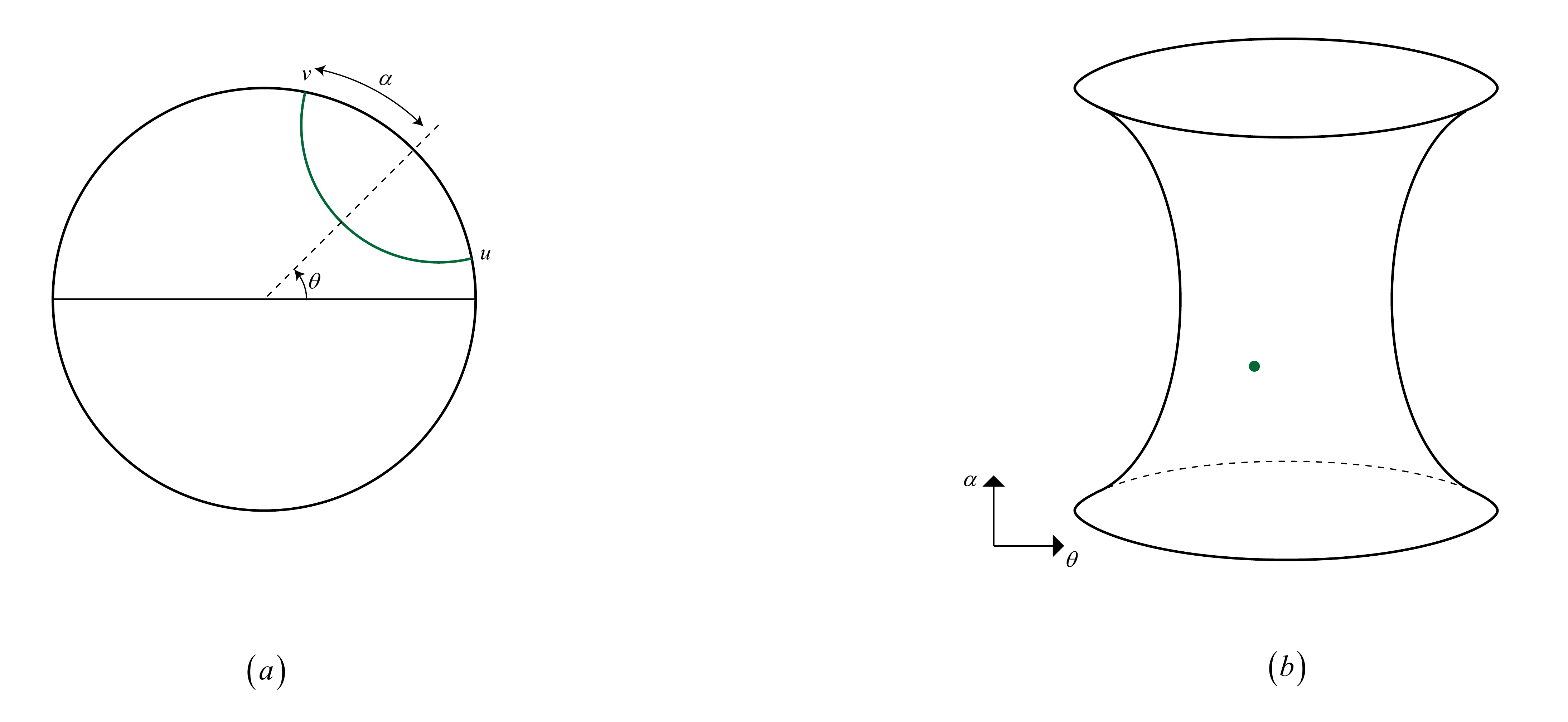}
		\caption{Geodesics (green lines) in the hyperbolic plane (left figure) are mapped to points on kinematic space (right figure).}
		\label{fig52}
	\end{center}	
\end{figure}

Our holographic qubit threads model also has a close connection to the so-called kinematic space. Kinematic space is an example of the application of quantum information theory to the holographic duality~\cite{Czech:2015kbp,Czech:2015qta}. Briefly speaking, it is a dual space wherein each point is one-to-one mapped from a pair of points on the boundary of the original geometry. 

\begin{figure}[htbp]     \begin{center}
		\includegraphics[height=7cm,clip]{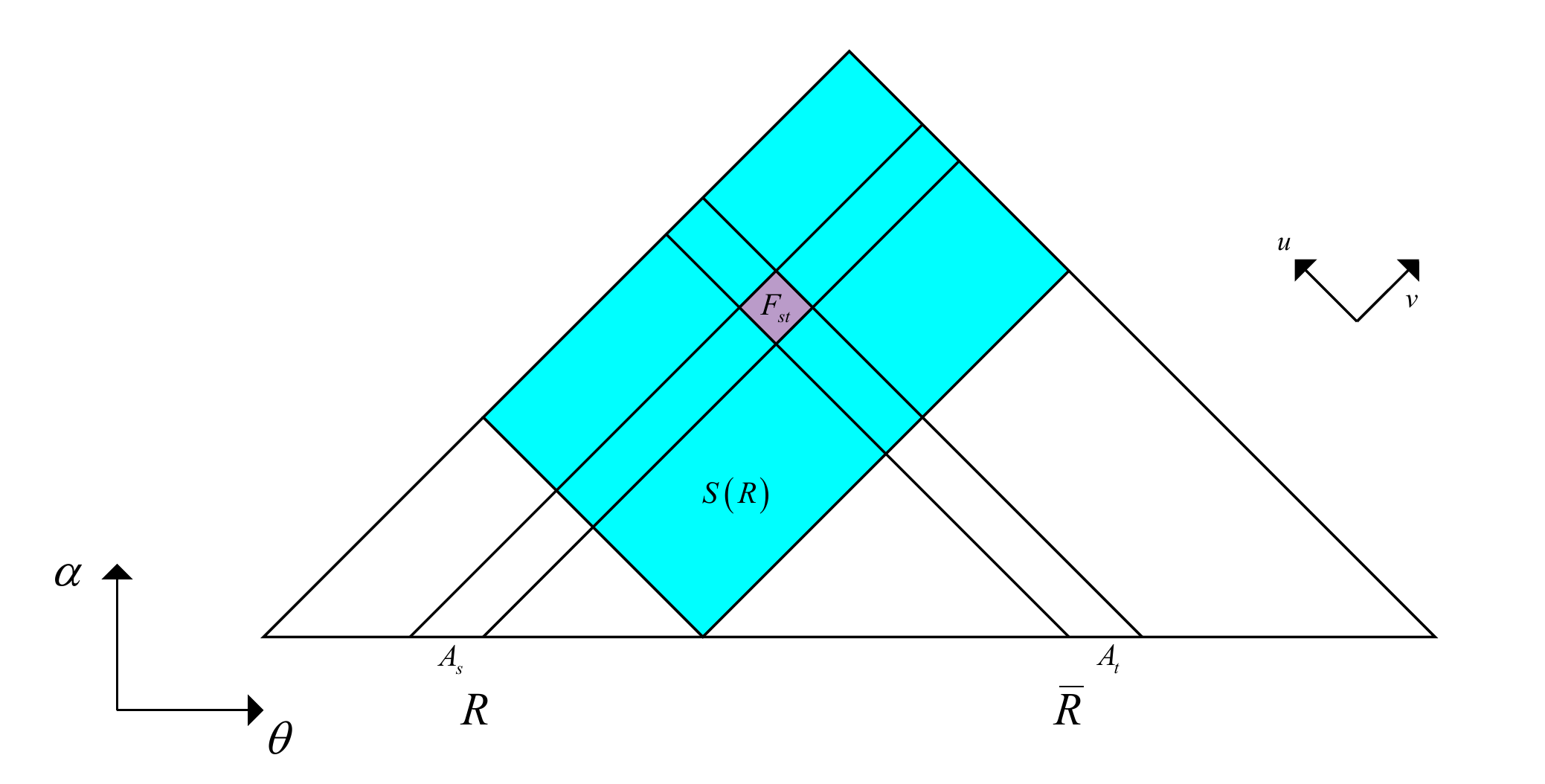}
		\caption{The entanglement entropy of the subregion $R$ in the boundary CFT can be represented as the volume of the causal diamond ${\diamondsuit _R}$ (in blue) in the kinematic space, while the flux ${F_{st}}$ of a thread bundle connecting two elementary regions ${A_s}$ and ${A_t}$ is precisely given by the volume of the diamond region ${\diamondsuit _{st}}$ (in mauve).		}
		\label{fig522}
	\end{center}	
\end{figure}

As shown in figure~\ref{fig52}, consider a pair of points parameterized by $\theta$, $\alpha$ on a time slice of the two-dimensional holographic CFT, which can also be parameterized by $u$, $v$, where $u = \theta  - \alpha$, $v = \theta  + \alpha$. Inspired by Crofton's formula in flat space,~\cite{Czech:2015kbp,Czech:2015qta} proposed that the geodesic $l(\theta ,\alpha )$ (or $l(u,v)$) connecting this pair of points can be bijective mapped to a point $(\theta ,\alpha )$ (or $(u,v)$) in a new two-dimensional space, which can then be equipped with a metric by using the generalized Crofton's formula, and go by the name of the holographic kinematic space. In the framework of holographic principle, since the RT formula relates the length of a geodesic in the bulk to the entanglement entropy of boundary CFT, this construction can also be described purely in the language of boundary CFT: The kinematic space is a dual space wherein each point $(\theta ,\alpha )$ (or $(u,v)$) is one-to-one mapped from a pair of points parameterized by $\theta$, $\alpha$ (or $u$, $v$) on the boundary of the original geometry, and its metric (i.e., its spatial volume density) is defined by the conditional mutual information (CMI)~\cite{Czech:2015kbp,Czech:2015qta},
\be\label{dv} d{V_{kin}} = \frac{1}{2}CMI,\ee
where
\be CMI(A,C\left| B \right.) = S\left( {AB} \right) + S(BC) - S(ABC) - S(B).\ee
In particular, for a time-slice of pure $Ad{S_3}$, the  metric  on the corresponding  kinematic  space  is exactly  the  two-dimensional  de  Sitter  metric~\cite{Czech:2014ppa}:
\be ds_{kin}^2 = \frac{L}{{{{\sin }^2}\alpha }}\left( { - d{\alpha ^2} + d{\theta ^2}} \right).\ee
It has been pointed out in~\cite{Czech:2015kbp,Czech:2015qta} that given the formula~(\ref{dv}), the entanglement entropy of a subregion $R$ in the boundary CFT can then be represented as the volume of a corresponding region ${\diamondsuit _R}$ (called causal diamond) in the kinematic space:
\be\label{sr}S(R) = {V_{kin}}{\rm{(}}{\diamondsuit _R}).\ee
As shown in figure~\ref{fig522}, ${\diamondsuit _R}$ is defined as a diamond-shaped region bounded by the light rays originating from the boundary points of $R$ (and its complement $\bar R$) in the kinematic space, which is represented as a blue region in the figure.

Now notice that in the holographic qubit threads model, the value of each thread bundle flux is given by the expression~(\ref{fij}). In other words, CMI and PEE, or CFF, are actually exactly the same thing that in a sense characterize the entanglement density. Naturally, since each thread in the holographic qubit threads model also relates a pair of points on the boundary, it can be bijective mapped to a point in the kinematic space. As a result, we can view each thread in our holographic qubit threads model as a point in a kinematic space, and vice versa.

The key point is that according to thread/state correspondence, now every point in the kinematic space is also considered as a qubit, namely a quantum superposition of two orthogonal basic states. Therefore, in fact, our thread/state rules can be regarded as indicating the microstates of the kinematic space. More specifically, as shown in figure~\ref{fig522}, in the holographic qubit threads model, the flux ${F_{st}}$ of a thread bundle connecting two elementary regions ${A_s}$ and ${A_t}$ (which are the internal regions in $R$ and $\bar R$ respectively) is precisely given by the volume of a diamond region ${\diamondsuit _{st}}$ in the kinematic space, which is defined as the region enclosed by the intersecting light rays emitted from the ends of ${A_s}$ and ${A_t}$ at the kinematic space boundary:
\be{F_{st}} = {V_{kin}}({\diamondsuit _{st}}).\ee
Now, we can give an interesting physical interpretation of (\ref{sr}). According to the thread/state rules, the points in the kinematic space can be regarded as qubits independent of each other. Denote the number of microstates of region ${\diamondsuit _R}$ as $\Omega ({\diamondsuit _R})$, and the number of microstates of region ${\diamondsuit _{st}}$ as $\Omega ({\diamondsuit _{st}})$, we have obviously
\be\Omega ({\diamondsuit _{st}}) = {2^{{F_{st}}}},\ee
and the von Neumann entropy of ${\diamondsuit _R}$ is equal to
\be S({\diamondsuit _R}) = \ln \Omega (R) = \ln \left( {{2^{\sum\limits_{s,t} {{F_{st}}} }}} \right) = \sum\limits_{s,t} {{F_{st}}}. \ee
by (\ref{fij}),
\be\sum\limits_{s,t} {{F_{st}}}  = \sum {CMI}, \ee
then by (\ref{dv}), one obtain (\ref{sr}).

In other words, by counting the number of microstates, one can naturally understand that the entropy is proportional to the volume in the kinematic space. It's interesting and instructive to look at this the other way around, as we can see, that a construction of microstates wherein the entropy is positively related to the volume (the characteristic of normal thermodynamic systems) actually can correspond to a construction wherein the entropy is positively related to the area (the characteristic of the holographic gravity system). Kinematic space is also closely related to the interpretation of the MERA tensor network~\cite{Czech:2015kbp,Czech:2015qta}, therefore, we expect that our thread/state correspondence will deepen the understanding of both.

A necessary comment: It is important to note that, in despite of the direct connection between the holographic qubit threads model and the kinematic space, there is actually one significant difference between the two: In fact, CMI threads that one-to-one map to points in the kinematic space do not need to meet the strict mathematical requirements (especially the density bound) of the bit threads; instead, these CMI threads only need to meet the constraints on the thread bundle fluxes (see~\cite{Lin:2021hqs} for a discussion of this issue). On the other hand, qubit threads that satisfy the non-trivial bulk density bound constraint of bit threads have the potential to extract more information in physics, although the technical difficulties will increase mathematically. For example, we can further consider the connection between the more general non-locking bit thread configurations and the more general bulk surfaces. Furthermore, various updated versions of bit threads have been very useful tools for studying various aspects of the relationship between geometry and quantum entanglement, it is possible to further consider how the thread/state rules adapt to the covariant bit threads~\cite{Headrick:2022nbe} of the covariant RT formula~\cite{Hubeny:2007xt}, the quantum bit threads~\cite{Agon:2021tia,Rolph:2021hgz} that can account for the bulk quantum corrections to the RT formula~\cite{Faulkner:2013ana,Engelhardt:2014gca}, the Lorentzian bit threads~\cite{Pedraza:2021fgp,Pedraza:2021mkh} that can characterize the holographic complexity~\cite{Brown:2015bva,Susskind:2014rva}, the hyperthreads~\cite{Harper:2021uuq,Harper:2022sky} that can study the multipartite entanglement, and so on, and may provide useful insights on all of these topics.

%%%%%%%%%%%%%%%%%%%%%%%%%%%%%%%%%%%%%%%%%%%%%%%%%%%%%%%%%%%%%%%%%%%%%%
\section{Bit threads and the connectivity of spacetime}\label{sec6}
%%%%%%%%%%%%%%%%%%%%%%%%%%%%%%%%%%%%%%%%%%%%%%%%%%%%%%%%%%%%%%%%%%%%%%

%%%%%%%%%%%%%%%%%%%%%%%%%%%%%%%%%%%%%%%%%%%%%%%%%%%%%%%%%%%%%%%%%%%%%%

\begin{figure}[htbp]     \begin{center}
		\includegraphics[height=8cm,clip]{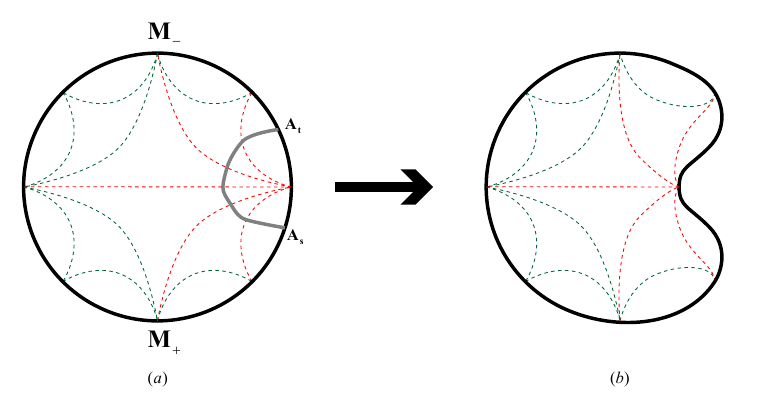}
		\caption{Removing the entanglement connecting a certain elementary region ${A_s}$ in ${M_ + }$ with another elementary region ${A_t}$ in ${M_ - }$, the area of the RT surfaces that intersect with thread bundle ${\vec v_{st}}$ will accordingly decrease by a value of ${F_{st}}$. The removed thread bundle is marked as a thick grey line, and the affected RT surfaces are marked as red dashed lines.}
		\label{fig8}
	\end{center}	
\end{figure}

Next, through a thought experiment similar to Raamsdonk's famous ``It from qubit" experiment~\cite{VanRaamsdonk:2010pw}, we are going to show how our qubit thread configuration model can describe the relationship between spacetime and the qubits distilled from the boundary quantum system in a sense. 

As shown in figure~\ref{fig8}, consider dividing the whole boundary quantum system $M$ into two parts, denoted as ${M_ + }$ and ${M_ - }$, each of which contains large amounts of elementary regions (for simplicity, we only schematically draw four fundamental regions in each part). Let us denote the elementary regions in ${M_ + }$ and ${M_ - }$ as ${A_s}$ and ${A_t}$ respectively. Now imagine that we remove the entanglement connecting a certain elementary region ${A_s}$ in ${M_ + }$ with another elementary region ${A_t}$ in ${M_ - }$. Then this implies that the flux ${F_{st}}$ of the qubit thread bundle ${\vec v_{st}}$ characterizing this part of the entanglement is changing from a certain value to zero before and after this operation. The point is that, as we have understood from our previous discussion, the thread bundle ${\vec v_{st}}$ not only contributes to the SS state of the RT surface that separates ${M_ + }$ and ${M_ - }$, but also contributes to the SS states of a series of RT surfaces it passes through. As shown in the figure, then, when this part of entanglement is taken away, in other words, when the flux ${F_{st}}$ of the thread bundle ${\vec v_{st}}$ becomes zero, the area of the RT surfaces that intersect with it will accordingly decrease by a value of ${F_{st}}$. Although we still do not fully figure out how to decode the geometric information of any surface or any region in the bulk from the entanglement information in the boundary CFT states completely, however, the area changes of these RT surfaces can quantitatively characterize how the bulk spacetime shrinks as the quantum entanglement is taken away to some extent. This is very interesting. This quantitative calculation can in principle be explicitly implemented by the following formula in the locking bit thread configuration~\cite{Lin:2021hqs}:
\be\begin{array}{l}
	{\rm{Are}}{{\rm{a}}_{a\left( {a + 1} \right) \ldots b}} = 4{G_N}\sum\limits_{s,t} {{F_{st}}} ,\\
	s \in \left\{ {a,a + 1, \cdots ,b} \right\},t \notin \left\{ {a,a + 1, \cdots ,b} \right\}
\end{array},\ee
where ${\rm{Are}}{{\rm{a}}_{a\left( {a + 1} \right) \ldots b}}$ represents the area of the RT surface associated with a connected composite region $R = {A_{a\left( {a + 1} \right) \cdots b}} \equiv {A_a} \cup {A_{a + 1}} \cup  \cdots  \cup {A_b}$.

\begin{figure}[htbp]     \begin{center}
		\includegraphics[height=12cm,clip]{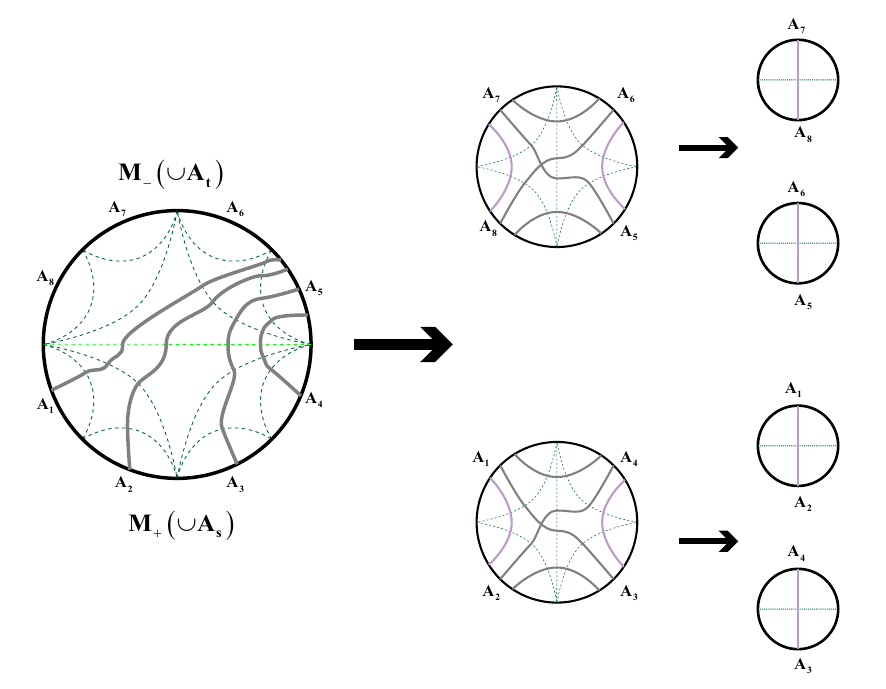}
		\caption{Removing all entanglement between the set of ${A_s}$ and the set of ${A_t}$ (we only draw a few of relevant thread bundles, i.e., the thick grey lines in the figure), the original bulk spacetime $N$ is split into two independent holographic bulk spacetimes ${N_ + }$ and ${N_ - }$. With this process repeating, finally, a bulk spacetime $N$ will eventually disintegrate into a tremendous number of small bubbles.}
		\label{fig9}
	\end{center}	
\end{figure}
As shown in figure~\ref{fig9}, let us further remove all entanglement between ${A_s}$ and ${A_t}$ such that all the fluxes ${F_{st}}$ of the thread bundles connecting ${A_s}$ and ${A_t}$ vanish, but the entanglement between the internal elementary regions within ${M_ + }$ and those within ${M_ - }$ is still preserved. Under such an operation, obviously, the area of the RT surface separating ${M_ + }$ and ${M_ - }$ degenerates to zero, which implies that the two parts ${M_ + }$ and ${M_ - }$ have been disconnected from each other. Thus, the original bulk spacetime $N$ is split into two independent holographic bulk spacetimes ${N_ + }$ and ${N_ - }$. Meanwhile, the remaining entanglement between the internal elementary regions of ${M_ + }$ and ${M_ - }$ can still maintain the internal spacetime geometry of ${N_ + }$ and ${N_ - }$ per se, which is reflected in the fact that the existing qubit threads can still continue to read out the SS states of the extremal surfaces inside ${N_ + }$ and ${N_ - }$. In principle，through the changes of these RT surfaces before and after removing the entanglement between ${M_ + }$ and ${M_ - }$, we can, in a sense, quantitatively analyze how the two new independent holographic bulk spacetimes ${N_ + }$ and ${N_ - }$ come into being from the original bulk $N$. Our qubit thread model thus visualizes the connection between the entanglement of the boundary quantum system and the connectivity of the holographic bulk spacetime. The process in figure~\ref{fig9} can be repeated indefinitely. One can further divide the boundaries of ${N_ + }$ and ${N_ - }$ into two halves, and then remove the entanglement between the two halves, so that each bulk will be further divided into two new smaller independent spacetimes. With this process repeating, finally, a bulk spacetime $N$ will eventually disintegrate into a tremendous number of small bubbles. Conversely, this implies that, one can build a continuous spacetime using quantum entanglement. Figuratively speaking, qubit threads can play the roles of ``sewing'' a spacetime. It extracts the entanglement information from the boundary quantum system, and then builds a spacetime just like sewing many small fragments into a sweater!

\begin{figure}[htbp]     \begin{center}
		\includegraphics[height=4.5cm,clip]{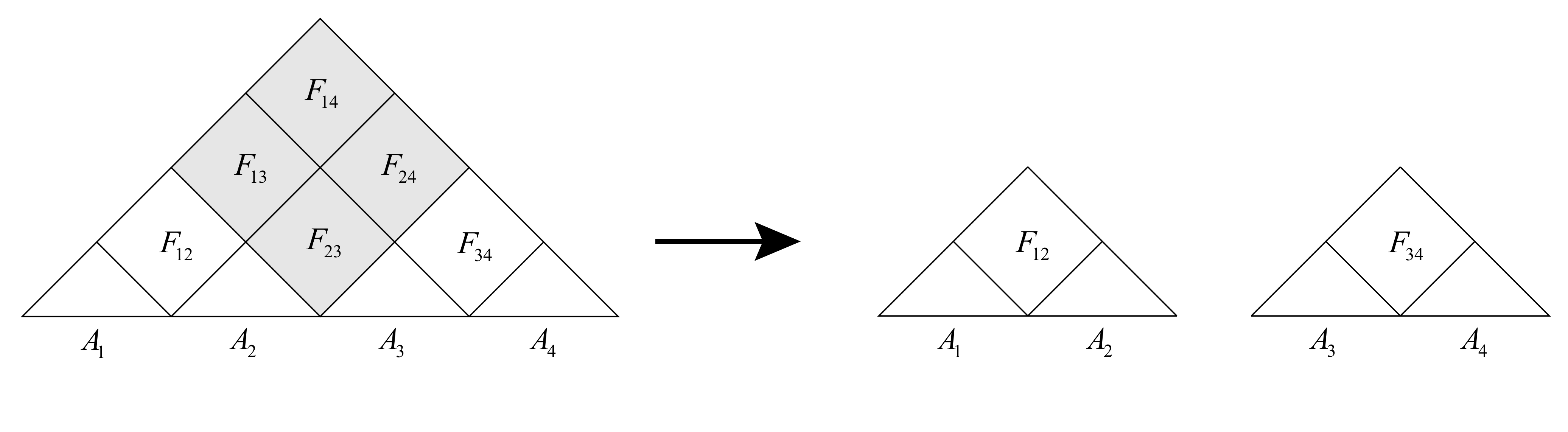}
		\caption{ Taking off the corresponding ``external threads'' entanglement corresponds to the removal of a diamond-like region in the kinematic space, and the original kinematic space was separated into two disconnected new kinematic spaces.}
		\label{fig62}
	\end{center}	
\end{figure}

The above thought experiment of splitting spacetime by removing the qubit thread bundles (or CMI, conditional mutual information) that represent quantum entanglement can also be described in the language of kinematic space. A schematic example in the viewpoint of the kinematic space that is dual to this process is shown in figure~\ref{fig62}. This simple case corresponds to the lower left of figure~\ref{fig9}, wherein the boundary quantum system is divided into four elementary regions: ${A_1}$, ${A_2}$, ${A_3}$, and ${A_4}$. Intuitively, taking off the corresponding ``external threads'' entanglement as shown in the figure corresponds to the removal of a diamond-like region in the kinematic space, which is the so-called casual diamond used to calculate the entanglement entropy between ${A_1}{A_2}$ and ${A_3}{A_4}$, and it turns out that the original kinematic space was separated into two disconnected new kinematic spaces! Furthermore, the two child kinematic spaces can nicely continue to describe the entanglement structure inside the two corresponding sub-spacetimes in figure~\ref{fig9}. In this dual viewpoint, this process can also be repeated all the way. We see that all the ideas are consistent.

%%%%%%%%%%%%%%%%%%%%%%%%%%%%%%%%%%%%%%%%%%%%%%%%%%%%%%%%%%%%%%%%%%%%%%
\section{Conclusions and discussions}\label{sec7}
%%%%%%%%%%%%%%%%%%%%%%%%%%%%%%%%%%%%%%%%%%%%%%%%%%%%%%%%%%%%%%%%%%%%%%
In this paper, we introduce a property of bit threads that has not been explicitly proposed before, which explicitly gives a physical interpretation of bit threads. Following the nomenclature of surface/state correspondence, we can refer to it as thread/state correspondence. More specifically, we propose that each bit thread can be endowed with two orthogonal states, which can be denoted as red state $\left| {{\rm{red}}} \right\rangle $ and blue state $\left| {{\rm{blue}}} \right\rangle $ respectively, and in a locking bit thread configuration, each bit thread is in a quantum superposition state of these two states. We show that such an interpretation of the bit threads leads to a clever and natural way of constructing the explicit expressions for the SS states corresponding to a set of bulk extremal surfaces in the SS duality, and nicely characterizing their entanglement structure. 

In this sense, we call them qubit threads and construct a holographic qubit threads model as an attempt of the new toy models of holographic principle, in which these qubit threads can also be understood as playing the role of CMI (conditional mutual information) threads. We explicitly show the connection between this model and the holographic tensor network models. In particular, we show how to understand the concept of ``disentanglers'' in the latter from the perspective of qubit threads. The relationship with the holographic kinematic space is also studied, that is, the flux of a thread bundle connecting two elementary regions can exactly match the volume of a corresponding diamond region bounded by light rays in the kinematic space. Then thread/state correspondence naturally  endows the kinematic space with the interpretation of microscopic states such that to explain that the entropy is proportional to volume therein. Finally, we show that our holographic qubit threads model can in some sense quantitatively characterize the famous ``It from qubit'' thought experiment~\cite{VanRaamsdonk:2010pw}, that is, by removing the entanglement in the boundary quantum system, the bulk spacetime will accordingly deform, or even break up.

In recent years, the idea of holographic tensor networks or quantum information theory has proved to be enlightening to the issue of spacetime emergence (``it from qubit''), while through the study of this paper, we have seen that the concept of (long-range) ``threads'' can provide a new perspective that is different from, but closely related to the local ``tensors'' (or quantum circuit gates). Although our thread/state correspondence rules for locking thread configurations merely apply to bulk extremal surfaces at present, it is a suggestive step towards the issue of spacetime emergence. Since the concept of bit threads originates from the duality of the optimization problem of the areas of geometric surfaces and the optimization problem of the fluxes of thread flows, we can further ask whether similar rules can be found and applied to the more general non-locking bit thread configurations so that one can further read the SS states of the more general bulk surfaces. In the longer term, an even more tantalizing idea is to further cast off the metric information of the background spacetime and completely reconstruct the bulk geometry only from the properties of the quantum state assigned to the bit threads.

Furthermore, actually there are various updated versions of bit threads, which are very useful tools for studying various aspects of the relationship between geometry and quantum entanglement, such as the covariant bit threads~\cite{Headrick:2022nbe}, the quantum bit threads~\cite{Agon:2021tia,Rolph:2021hgz}, the Lorentzian bit threads~\cite{Pedraza:2021fgp,Pedraza:2021mkh}, the hyperthreads~\cite{Harper:2021uuq,Harper:2022sky}, etc. It is likely that our thread/state correspondence rules could be further adapt to these similar objects and lead to deeper or clearer understandings of these different important aspects of the holographic duality.

%%%%%%%%%%%%%%%%%%%%%%%%%%%%%%%%%%%%%%%%%%%%%%%%%%%%%%%%%%%%%%%%%%%%%%
\section*{Acknowledgement}
%%%%%%%%%%%%%%%%%%%%%%%%%%%%%%%%%%%%%%%%%%%%%%%%%%%%%%%%%%%%%%%%%%%%%%
We would like to thank Ling-Yan Hung and Yuan Sun for useful discussions.

%%%%%%%%%%%%%%%%%%%%%%%%%%%%%%%%%%%%%%%%%%%%%%%%%%%%%%%%%%%%%%%%%%%%%%
\begin{appendix}
%%%%%%%%%%%%%%%%%%%%%%%%%%%%%%%%%%%%%%%%%%%%%%%%%%%%%%%%%%%%%%%%%%%%%%

%%%%%%%%%%%%%%%%%%%%%%%%%%%%%%%%%%%%%%%%%%%%%%%%%%%%%%%%%%%%%%%%%%%%%%
\section{proof based on strong duality of convex program}\label{app}
$\bf{Statement}$: For a (Riemannian) manifold $C$ whose boundary $\partial C$ consisting of three minimal surfaces (denoted as elementary regions $1$, $2$, $3$ respectively), there always exists a multiflow $V$ consisting of three component flows ${\vec v_{12}}$, ${\vec v_{13}}$, ${\vec v_{23}}$ (we define only ${\vec v_{ij}}$ with $i<j$), which by definition satisfies: 
\be\label{a1}\hat n \cdot {\vec v_{ij}}\left| {_k} \right. = 0\quad \left( {k \ne i,j} \right),\ee
\be\label{a2}\nabla  \cdot {\vec v_{ij}} = 0,\ee
\be\label{a3}\left| {{{\vec v}_{ij}}} \right| \le \frac{1}{{4{G_N}}},\ee
that can lock regions $1$, $2$ and $3$ simultaneously and meet the addition boundary conditions:
\be\label{a4}{\left. {\left| {{{\vec v}_{ij}}} \right|} \right|_i} = {c_{ij}},\;{\left. {\left| {{{\vec v}_{ij}}} \right|} \right|_j} = {c'_{ij}},\ee
i.e,, the densities of the three component flows on the boundary must be constant.

$\bf{Proof}$: The proof is almost exactly the same as presented in~\cite{Headrick:2020gyq}.

Firstly, due to the max flow-min cut theorem~\cite{Headrick:2017ucz}, for any multiflow (not necessarily locking) that meets the above conditions, we immediately have~\cite{Headrick:2020gyq,Cui:2018dyq}
\be\int_1 {\left( {{{\vec v}_{12}} + {{\vec v}_{13}}} \right) + \int_2 {\left( {{{\vec v}_{12}} + \;{{\vec v}_{23}}} \right)} }  + \int_3 {\left( {{{\vec v}_{13}} + {{\vec v}_{23}}} \right)}  \le {S_1} + {S_2} + {S_3},\ee
or by the divergenceless property (\ref{a2}), i.e.,
\be\label{sma}\sum\limits_{i < j} {2\int_i {{{\vec v}_{ij}}} }  \le \sum\limits_i {{S_i}} .\ee
To prove the existence of such a locking configuration, the strategy is to use the convex duality technique from the theory of convex program~\footnote{ See~\cite{Cui:2018dyq,Headrick:2017ucz,Headrick:2020gyq} for a physicist-friendly introduction to the technical details on this topic.
} to prove the existence of a feasible multiflow such that
\be\sum\limits_{i < j} {2\int_i {{{\vec v}_{ij}}} }  \ge \sum\limits_i {{S_i}}. \ee
Combined with (\ref{sma}), this multiflow must be a locking bit thread configuration that locks $1$, $2$, and $3$ simultaneously.

Thus, we just need to deal with the following $primal$ program:
Given conditions (\ref{a1})-(\ref{a4}), over all feasible multiflows ${{{\vec v}_{ij}}}$, maximize the objective function:
\be F = \sum\limits_{i < j} {2\int_i {{{\vec v}_{ij}}} }. \ee
Following~\cite{Cui:2018dyq,Headrick:2017ucz,Headrick:2020gyq}, since the equality constraints (\ref{a1})(\ref{a2})(\ref{a4}) are affine, and the inequality constraint (\ref{a3}) is convex, this problem is a convex program, then we can use strong duality of convex programs, which indicates that under certain mild conditions (so-called Slater’s condition) the dual program has the same optimal value as the primal.
We will choose the boundary condition (\ref{a1}) and (\ref{a4}) to be $implicit$ and the divergenceless constraint (\ref{a2}) and norm bound (\ref{a3}) to be $explicit$. This means that we introduce Lagrange multipliers, ${\psi _{ij}}$ ($i < j$) and $\lambda$ (subject to $\lambda  \ge 0$) for constraints (\ref{a2}), (\ref{a3}) respectively. We then solve the maximization problem without imposing these two constraints, but while still imposing constraints (\ref{a1}) and (\ref{a4}). Adding the Lagrange multiplier terms to the objective, we arrive at the following Lagrangian functional:
\be L = \sum\limits_{i < j} {2\int_i {{{\vec v}_{ij}}} }  + \int_C {\sqrt g \left( {\sum\limits_{i < j} {{\psi _{ij}}\nabla  \cdot {{\vec v}_{ij}}}  + \lambda \left( {\frac{1}{{4{G_N}}} - \left| {{{\vec v}_{ij}}} \right|} \right)} \right)}. \ee
Integrating  the  divergence  term  by  parts and rewriting it slightly, we obtain
\be\label{l2} L = \sum\limits_{i < j} {\left( {\int_i {{{\vec v}_{ij}}\left( {2 - {\psi _{ij}}} \right)}  - \int_j {{{\vec v}_{ij}}{\psi _{ij}}} } \right)}  + \int_C {\sqrt g \left( {\frac{\lambda }{{4{G_N}}} - \sum\limits_{i < j} {\left( {\nabla {\psi _{ij}} \cdot {{\vec v}_{ij}} + \lambda \left| {{{\vec v}_{ij}}} \right|} \right)} } \right)}. \ee
We now maximize the Lagrangian with respect to ${\vec v_{ij}}$ (remember only to impose implicit constraints (\ref{a1}), (\ref{a4}) but not (\ref{a2}), (\ref{a4})). The according to the theory of convex duality, the requirement that the maximum is finite leads to constraints on the dual variables $\left\{ {{\psi _{ij}}} \right\}$, $\lambda$. 
Under the constraints (\ref{a1}) (\ref{a4}), in order for the maximum of first term of (\ref{l2}) to be finite with respect to ${\vec v_{ij}}$, the coefficient in front of ${\vec v_{ij}}$ must vanish, which leads to the constraints
\be{\left. {{\psi _{ij}}} \right|_i} = 2,\quad {\left. {{\psi _{ij}}} \right|_j} = 0,\ee
similarly, in the bulk, in order for the second term of (\ref{l2}) to have a finite maximum, there must be
\be\lambda  \ge \nabla {\psi _{ij}}.\ee
(Note that as a result, the constraint $\lambda  \ge 0$ is automatically satisfied and thus can be dropped.)
All in all, finally we  are left  with the  following  dual  program: 

Minimize $G = \frac{1}{{4{G_N}}}\int_C {\sqrt g \lambda }$ with respect  to $\left\{ {{\psi _{ij}}} \right\}$, $\lambda$, 

subject  to $\lambda  \ge \nabla {\psi _{ij}}$, ${\left. {{\psi _{ij}}} \right|_i} = 2$, ${\left. {{\psi _{ij}}} \right|_j} = 0.$

We note that in Section 3.1 of~\cite{Headrick:2020gyq} (see also~\cite{Cui:2018dyq}) the $infimum$ ${d^ * }$ for a program of the same form as the one here (the objective function is only up to a trivial $\frac{1}{{4{G_N}}}$ factor) has been computed directly, so we give the result directly:
\be\frac{1}{{4{G_N}}}\int_C {\sqrt g \lambda }  \ge \frac{1}{{4{G_N}}}\sum\limits_i {Area\left( i \right)}  \equiv \sum\limits_i {{S_i}}. \ee

Strong duality follows from the fact that Slater's condition is obeyed. Slater’s condition states that there exists a value for the primal variables such that all equality constraints are satisfied and all inequality constraints are strictly satisfied. This is the case here: the configuration ${\vec v_{ij}} = 0$ satisfies all the equality constraints (\ref{a1}) (\ref{a2}) (\ref{a4}) and strictly satisfies the norm bound (\ref{a3}). Therefore, strong duality holds:   the primal  and  dual programs  have  the  same  optimal  values, i.e., the $supremum$ ${p^ * }$ of the $primal$ program is equal to the $infimum$ ${d^ * }$ of the $dual$ program:
\be{p^ * } = {d^ * },\ee
that is,
\be{p^ * } \ge \sum\limits_i {{S_i}}. \ee
Therefore there must exist a bit thread configuration satisfies
\be\sum\limits_{i < j} {2\int_i {{{\vec v}_{ij}}} }  \ge \sum\limits_i {{S_i}},\ee
and we complete the proof.

\end{appendix}

%%%%%%%%%%%%%%%%%%%%%%%%%%%%%%%%%%%%%%%%%%%%%%%%%%%%%%%%%%%%%%%%%%%%%%

%\end{thebibliography}

\end{document}